\documentclass[12pt]{article}
\pdfoutput=1

\usepackage{paper2e}
\usepackage{mydefs2e}
\usepackage{xspace}
\usepackage{graphicx}
\usepackage{amssymb}

\usepackage{color}

\newcommand{\gap}{\hspace{0.05em}}

\newcommand{\Sla}[1]%
{\kern0.12em{\raise.15ex\hbox{$/$}\kern-.74em #1}}

\renewcommand{\d}{\partial}

\newcommand{\beqa}{\begin{eqnarray}}
\newcommand{\eeqa}{\end{eqnarray}}

\begin{document}

\begin{titlepage}

\title{The $a$-theorem and the\\\medskip
Asymptotics of 4D Quantum Field Theory}

\author{Markus A. Luty}

\address{Institut f\"ur Theoretische Physik der Universit\"at Heidelberg\\
Heidelberg, Germany}

\address{Physics Department, University of California Davis\\
Davis, California 95616\,\footnote{Permanant address}}

\author{Joseph Polchinski}

\address{Kavli Institute for Theoretical Physics,
University of California Santa Barbara\\
Santa Barbara, California 93106}

\author{Riccardo Rattazzi}

\address{Institut de Th\'eorie des Ph\'enom\`enes Physiques, EPFL\\
Lausanne, Switzerland}

\begin{abstract}

\end{abstract}
We study the possible IR and UV asymptotics of 4D
Lorentz invariant unitary quantum field theory.
Our main tool is a generalization of the Komargodski-Schwimmer proof
for the $a$-theorem.
We use this to rule out a large class of
renormalization group flows that do not asymptote to conformal
field theories in the UV and IR.
We show that the only possible UV and IR asymptotics described by
perturbation theory have a vanishing trace of the stress-energy tensor, 
and are therefore conformal.
Our arguments hold even for theories with gravitational anomalies.
We also give a non-perturbative argument that excludes theories with scale
but not conformal invariance.
This argument holds for theories in which the stress-energy tensor
is sufficiently nontrivial in a technical sense that we make precise.
\end{titlepage}

\section{Introduction}
\label{sec:intro}
In this paper we study the 
4D Lorentz invariant, unitary quantum
field theory in the asymptotic UV and IR limits.
In every 4D theory where these asymptotics are known,
they are described by a conformal field theory (CFT).
For example, QCD in the chiral limit
approaches a free theory of massless vector
particles in the far UV, and a theory of free massless scalars
(pions) in the IR.
If we add quark masses, the pions get a mass and the IR
theory is trivial.
A more complicated example is QCD with $N_f \simeq \frac{11}{2} N_c$,
which asymptotes to an interacting conformal field theory in the IR.
In many cases, the UV limit is not well-defined,
and we interpret these theories as effective field theories
needing UV completion.
Given the wealth of possible UV theories
(provided by string theory, for example)
it seems unlikely that we can make any definite classification
of UV limits of quantum field theories.
For the IR limits, it is a reasonable conjecture that all such
theories are conformal field theories (or trivial).

A useful way of approaching this question is Wilson's
renormalization group (RG) flow in the space of theories.
The IR asymptotics is then in one-to-one correspondence with
the IR behavior of the RG flow.
Theories like QCD flow from one fixed point to another, 
as illustrated in Fig.~1a.
Other more exotic possibilities are limit cycles (Fig.~1b)
or ergodic behavior (Fig.~1c).

In this paper, we report on progress in ruling out
RG flows that do not asymptote to
CFTs in the UV or IR.
We demonstrate that within perturbation theory, all theories
that remain perturbative in the UV or IR asymptote to a
CFT.
More precisely, we show that
\beq
T  \to 0
\eeq
as an operator statement, where $T = T^\mu{}_\mu$.
This means that correlation functions with one insertion of
$T$ with arbitrary numbers of elementary fields
asymptote to zero.
In particular, this excludes perturbative 4D theories with scale
but not conformal invariance (SFTs).

\begin{figure}[t]
\begin{center}
\includegraphics[scale=0.85]{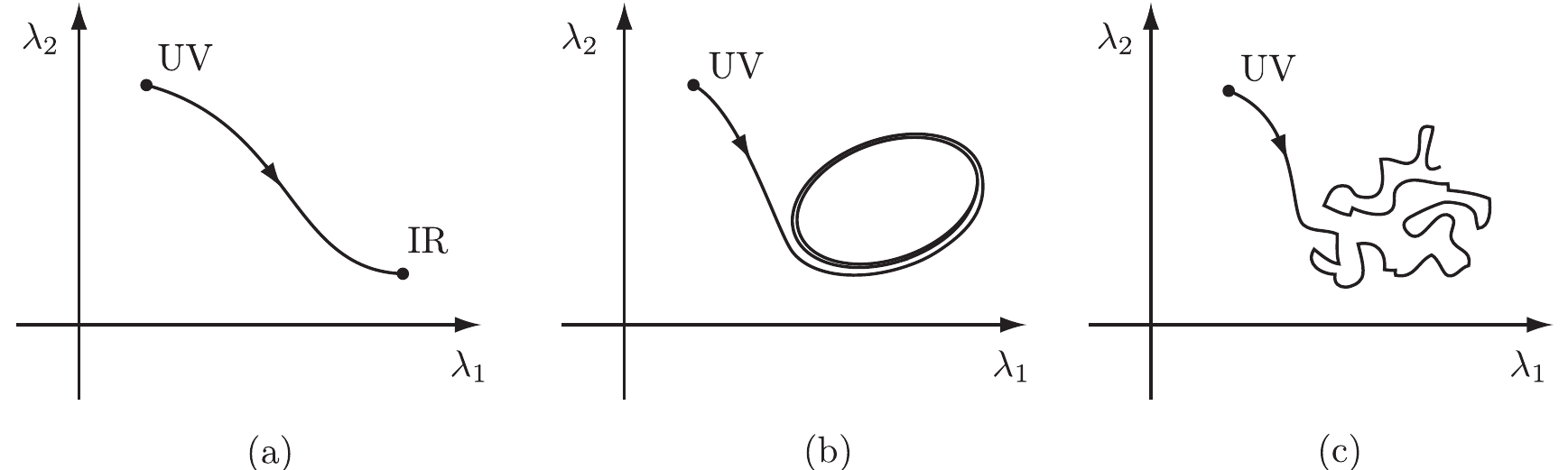} 
\end{center}
\caption{{\it A priori} possible IR behavior of renormalization group flows.}
\end{figure} 

Beyond perturbation theory, we show that SFTs
have a stress-energy tensor whose trace $T$
is almost trivial, in a technical
sense that we make precise below.
We believe that this implies that $T \equiv 0$ as
an operator statement, but are unable to give a rigorous proof.

The main tool in establishing these results
is a generalization of the
recent proof of the $a$-theorem
by Komargodski and Schwimmer (KS)~\cite{KS,Ka}.
This asserts that for theories that asymptote to CFTs both
in the UV and the IR,
\beq
a_{\rm UV} \ge a_{\rm IR},
\eeq
where $a$ is the anomaly coefficient of the CFT
that describes the UV or IR limit.
We will give a version of
their proof that closes some potential loopholes in the original
argument in \Ref{KS}.
We emphasize however that the key points of our proof
are identical to the KS argument.

The idea is to consider the quantum field theory of interest
in a conformally flat metric of the form $e^{-2\tau(x)} \eta_{\mu\nu}$.
The effective action $W[\tau]$ then defines the matrix
elements of $T$ in flat spacetime.
Alternatively, we can view $W[\tau]$ as the action for
dilaton self-interactions obtained by integrating out
the quantum field theory.
This physical picture is not necessary for the argument,
but it makes the arguments clearer.
Following \Ref{KS} we define a particular on-shell forward
dilaton-dilaton scattering amplitude $A(s)$ from $W[\tau]$ 
that has no relevant or marginal counterterms.
We write
\beq[alphadefn]
A(s) = \frac{\al(s) s^2}{f^4},
\eeq
where $s$ is the square of the center-of-mass energy,
$f$ is the dilaton decay constant that counts powers of the dilaton field,
and $\al(s)$ is a dimensionless function of $s$.
The UV and IR limits of this scattering amplitude are completely
determined by the ``$a$'' conformal anomaly,
in the sense that
\beq[aUVIR]
\al(s \to \infty) - \al(s \to 0) = -8 \left(
a_{\rm UV} - a_{\rm IR} \right).
\eeq
This immediately relates $a_{\rm UV} - a_{\rm IR}$ to
the dilaton-dilaton scattering amplitude.
The \lhs of \Eq{aUVIR} can be shown to be positive in unitary theories
by a dispersive argument, thus proving the $a$-theorem.
Our discussion pays particular attention to the convergence of the
dispersion relation, which is crucial for the argument.

To find the restrictions on general perturbative flows, we use the 
following logic.
We define $W[\tau]$ and the dilaton-dilaton scattering amplitude 
as in the $a$-theorem argument.
In a perturbative theory, $\al(s)$ is given by a power series in renormalized
couplings renormalized at the scale $s$, with no counterterm.
Therefore $\al(s)$ is bounded at all scales where perturbation
theory is valid.  The contour argument of Komargodski and Schwimmer is then adapted to show that the beta functions (defined as the coefficients appearing in the expansion of $T$ in local operators) must vanish in the UV and IR limits.

For non-perturbative SFTs, we use the anomalous Ward identities of
scale invariance to demonstrate that $\al(s)$ is exactly constant.
This has no imaginary part, and for unitary theories this implies that
\beq[Ttrivial]
\bra{X} T(p_1) T(p_2) + T(p_1 + p_2) \ket{0} \to 0
\eeq
for all states $\ket{X}$.
In a perturbative theory, we show that this can hold only if
$T \to 0$ as an operator.
We cannot rigorously prove that this implies $T \equiv 0$ in general,
but we give some reasons for thinking that this is the case.
We test these ideas by showing that the imaginary part of the
amplitude vanishes in the 4D Riva-Cardy model, a non-unitary SFT.

Non-perturbative SFTs that can be deformed to a CFT in the UV
(IR) at an adjustable scale $\La_{\rm UV}$ ($\La_{\rm IR}$) are
ruled out by our arguments.
In such theories $\al(s)$ diverges at the UV (IR)
as $\La_{\rm UV} \to \infty$ ($\La_{\rm IR} \to 0$),
which is incompatible with the fact that the anomaly
coefficient $a_{\rm UV}$ ($a_{\rm IR}$) of the UV (IR) CFT
is finite (see \Eq{aUVIR}).
This gives additional reasons for thinking that SFTs
are impossible in general.

The arguments we use depend heavily on the consistency
of the theory in a background metric.
Nonetheless, we show that the arguments hold even for theories
with gravitational anomalies because
the anomaly does not depend on the dilaton mode.
We do not consider theories that do not have a conserved stress-energy
tensor, which may for example emerge as low-energy effective theories
\eg\ from lattice models.

This paper is organized as follows.
In \S 2 we present the proof of the $a$-theorem.
In \S 3 we derive the restrictions on the renormalization
group flows of perturbative theories.
In \S 4 we discuss possible non-perturbative
theories with scale but not conformal invariance,
and we summarize in \S 5.
Some technical details of scale without conformal invariance
in perturbation theory in an appendix.

An explicit example of a perturbative SFT was recently proposed by
Fortin, Grinstein, and Stergiou in \Ref{FGSexample}.
However, it has now been 
recognized that this theory is exactly conformal~\cite{FGSta}.  
Our argument does not exclude SFTs in $4 + \ep$ dimensions, possible
examples of which were previously presented by the same authors 
\cite{FGS4minusepsilon}.
We are grateful to the authors of these papers
for extensive discussions of the previous 
version of our work, which have led us to a much clearer presentation.
In the previous version we identified the dilaton scattering amplitude with the running coefficient 
of the WZ term.  This is incorrect~\cite{FGSJO}, 
but this identification does not actually enter
into our argument --- see the discussion of \Eq{alrun}.  
These discussions have also helped us to understand the classic papers
\Refs{Osborn:1989td,Jack:1990eb,Osborn:1991gm}, enabling us to 
get a better understanding of the perturbative result,
as well as providing an alternative derivation of our results for 
the perturbative case. 

\section{The $a$-Theorem}
\label{sec:atheorem}
In this section we will present a proof of the $a$-theorem, 
filling in some details in the argument of \Refs{KS,Ka}.
Several of the steps of the
proof will be used in the generalizations that follow.

\subsection{The Dilaton as External Field
\label{sec:dilatonUVcouplings}}

We introduce the dilaton as the conformal mode of the
metric $g_{\mu\nu}$ by considering the theory in the background metric
\beq
\hat{g}_{\mu\nu} = e^{-2\tau} g_{\mu\nu}.
\eeq
We will eventually take $g_{\mu\nu} = \eta_{\mu\nu}$,
but we keep it general to make the general covariance of our
results explicit.

We consider a UV CFT deformed by relevant operators, generating
a nontrivial RG flow in the IR.
(We will not discuss flows induced by turning on moduli fields,
\eg in supersymmetric theories.)
The case of marginally relevant operators generating logarithmic flows
will also be included below.
The action is then
\beq[UVcoup]
S = S_{\rm UV} + \myint d^4 x \sqrt{-\hat{g}}
\sum_i c_i m^{4 - \De_i} \hat{\scr{O}}_i \,,
\eeq
where $S_{\rm UV}$ is the action of the UV CFT,
$\hat{\scr{O}}_i$ are relevant primary operators with dimension 
$\De_i < 4$,
$m$ is the mass scale associated with the flow, and
$c_i$ are dimensionless coefficients.
(Descendant operators are total derivatives, and therefore do not
deform the theory.)
We do not include irrelevant operators in the action because
this would change the UV behavior of the theory.
In other words, we are softly breaking the conformal symmetry
of the UV CFT.
In the presence of the perturbation above, the UV behavior of the
theory is governed by the UV CFT,
but there is a nontrivial flow at the scale $m$ where conformal
invariance is explicitly broken.
In this section we will assume that in the IR the theory
flows to a different IR CFT;
we will consider more general IR behavior in the following
sections.

Conformal transformations are the subgroup
of $\mbox{Weyl} \,\times\, \mbox{diffeormorphisms}$
that leave the flat space metric invariant, where
Weyl transformations are defined by%
\footnote{We follow the conventions of KS for the 
metric and dilaton.}
\beq[hatWeyl]
\hat{g}_{\mu\nu} \mapsto e^{2\si} \hat{g}_{\mu\nu},
\qquad
\hat{\scr{O}} \mapsto e^{-\si \De} \hat{\scr{O}},
\eeq
with $\si$ a general function of $x$.

We now consider the theory in the background metric 
$\hat{g}_{\mu\nu}= e^{-2\tau} g_{\mu\nu}$
introduced above.
The field $\tau$ is a redundant variable, since it can be
eliminated by the trivial gauge invariance
\beq[reparam]
\tau \mapsto \tau + \al,
\qquad
g_{\mu\nu} \mapsto e^{2\al} g_{\mu\nu}.
\eeq
This transformation acts trivially on the CFT fields.
The existence of the field $\tau$ allows us to define Weyl 
transformations \Eq{hatWeyl} as acting on the dilaton and leaving
$g_{\mu\nu}$ invariant:
\beq[Weylnew]
\tau \mapsto \tau - \si,
\qquad
g_{\mu\nu} \mapsto g_{\mu\nu},
\qquad
\hat \scr{O} \mapsto e^{-\si \De} \hat \scr{O}.
\eeq
Because of the gauge invariance \Eq{reparam}
the fields $\tau$ and $g_{\mu\nu}$ are not uniquely defined,
but we will simply choose canonical background field configurations
$g_{\mu\nu}$ and $\tau$.
For our main applications, $g_{\mu\nu} = \eta_{\mu\nu}$ so that
we are considering a conformally flat background.
Correlation functions of $\tau$ then define dilaton correlation
functions in flat spacetime, which will be the main object of
study in the following.

The UV theory is invariant under Weyl transformations up to the
Weyl anomaly~\cite{Duff}, so we have
\beq
\bal
\!\!\!\!\!\!\!
W[\hat{g}_{\mu\nu}] &= W_{\rm UV}[g_{\mu\nu}]
- \myint d^4 x \sqrt{-g} \, \tau
\left[ -a_{\rm UV} E_4(g) + c_{\rm UV} W^2(g) \right]
+ O(\tau^2)
\\
&\qquad\quad
{}+ \mbox{relevant terms}.
\eal
\eeq
The terms higher order in $\tau$ complete themselves into the
WZ term:
\beq[decouplingUV]
W[\hat{g}_{\mu\nu}] = W_{\rm UV}[g_{\mu\nu}] 
- S_{\rm WZ}[g_{\mu\nu}, \tau; a_{\rm UV}, c_{\rm UV}]
+ \mbox{relevant terms}.
\eeq
The WZ term is given by%
\footnote{We will be interested in the WZ term in flat spacetime
as a function of the dilaton field.
It is therefore worth noting that in flat spacetime
the WZ term is the unique term in
the dilaton Lagrangian that is invariant under special conformal
transformations only up to a total derivative term.
(A classification of dilaton invariants was given in \Ref{Nicolis:2008in}.)
The other terms in the dilaton effective action can be written in
terms of $\hat{g}_{\mu\nu}$ and are therefore exactly invariant.
This is analogous to the anomaly in the chiral Lagrangian for a
$G/H$ coset, where the WZ term is the unique term that shifts by a 
total derivative under $G$ transformations.}
\beq[WZ]
S_{\rm WZ}[g_{\mu\nu}, \tau; a, c] = \myint d^4 x \sqrt{-g}\,
\Bigl\{ & {-a} \Bigl[
\tau E_4(g) 
\nonumber\\
&\qquad\quad
{}+ 4 \bigl( R^{\mu\nu}(g) - \sfrac 12 g^{\mu\nu} R(g) \bigr) 
\Om^{-2} \d_\mu\Om \d_\nu\Om
\nonumber\\
&\qquad\quad
{}+4 \Om^{-3} (\d\Om)^2 \Box\Om 
-2  \Om^{-4} (\d\Om)^4 \Bigr]
\nonumber\\
&\ 
{}+ c \tau W^2(g) 
\Bigr\}.
\eeq
where 
\beq[Omegadef]
\Om = e^{-\tau}.
\eeq
The reason that the full WZ term appears in \Eq{decouplingUV}
is that it is the unique term that correctly reproduces 
the abelian nature of conformal transformations.
In the present context, this means that the
\rhs\ of \Eq{decouplingUV} is invariant under the
gauge transformations \Eq{reparam}:
\beq
\de S_{\rm WZ}[g_{\mu\nu}, \tau; a_{\rm UV}, c_{\rm UV}]
&= \myint d^4 x \sqrt{-g} \, \al \left[ -a_{\rm UV} E_4(g)
+ c_{\rm UV} W^2(g) \right]
\nonumber\\
&= \de W[g_{\mu\nu}].
\eeq

\Eq{decouplingUV} is basis for all the results in this paper.
It shows that the dependence on $\tau$ in the UV
comes entirely through the WZ term.
Note that even in flat space, the WZ term contains
dilaton self-interaction terms, which will be crucial
for the argument.
The relevant deformation terms depend on $\tau$,
but these are unimportant in the UV.

Note that if the IR theory is given by a CFT with only 
irrelevant deformations, exactly the same logic also gives
\beq[IRdilatondecouple]
W[\hat{g}_{\mu\nu}] = W_{\rm IR}[g_{\mu\nu}]
- S_{\rm WZ}[g_{\mu\nu}, \tau; a_{\rm IR}, c_{\rm IR}]
+ \mbox{irrelevant terms},
\eeq
where $W_{\rm IR}$ is the effective action for the IR CFT
with anomaly coefficients $a_{\rm IR}$ and $c_{\rm IR}$.
We will show below that in general the IR CFT contains 
relevant deformation terms that give a nonlocal relevant
correction to the \rhs\ of \Eq{IRdilatondecouple}.
However, we will show that $a_{\rm IR}$ is
nonetheless directly related to the on-shell
dilaton scattering amplitude, the crucial observable in the KS
argument.

We now return to the UV theory and 
consider the dilaton couplings arising from the relevant
deformation terms.
The original action is written in terms of CFT fields
$\hat\Phi$ and background metric $\hat{g}_{\mu\nu}$.
Defining
\beq
\hat{g}_{\mu\nu} = g_{\mu\nu} e^{-2\tau},
\qquad
\hat\Phi = e^{\tau \De_\Phi} \Phi
\eeq
the soft terms in the action have the form
\beq
S_{\rm soft}[\hat\Phi, \hat{g}_{\mu\nu}]
= \myint d^4 x\, \sqrt{-g} \, (m \Om)^{4 - \De} \scr{O}[\Phi,g_{\mu\nu}].
\eeq
We see that if we change variables to make decoupling manifest,
we automatically nonlinearly realize the Weyl invariance
in terms of $g_{\mu\nu}$ and $\tau$.
This gives a new perspective on the nonlinear realization of
conformal symmetry used in the KS argument.

\subsection{Dilaton Effective Field Theory
\label{sec:dilatoncouplings}}
In this section we are assuming that the IR dynamics is controlled
by a different CFT, so 
we can apply the same steps to the IR theory.
The IR theory is to be regarded as an effective field theory,
and we must include all couplings allowed by symmetries.
We are particularly concerned about possible relevant terms,
since these dominate in the IR.%
\footnote{For the UV theory we would also like to know the most
general relevant deformations allowed by symmetries in order to
study the most general theory that flows to the UV CFT in the UV.
The analysis below is relevant to this case as well.}

The observable we will be interested in is dilaton-dilaton
scattering in flat spacetime, in a low-energy expansion.
Specifically, we define the physical dilaton field $\varphi$ by
\beq[dildef]
\Om = 1 + \frac{\varphi}{f},
\eeq
where $f$ is the dilaton decay constant.
The WZ term \Eq{WZ} contains $O(E^4)$ dilaton self-couplings.
We want to see if there are other terms in the effective
theory that are more important in the IR.
We will show that the interactions in the WZ term dominate
at low energies provided that we impose the on-shell condition
$\Box\varphi = 0$, or
\beq[onshell]
\Box\Om = 0
\eeq
on the external dilaton lines, and fine-tune the IR cosmological
constant.
The cubic dilaton interaction in the WZ term
then vanishes, but the quartic term remains.
There are other possible definitions of the dilaton field
and on-shell condition,
but we will show that this choice is the one that ensures that
the WZ term dominates the amplitude.

In the IR effective theory $m$ is a UV cutoff scale,
and an operator $\scr{O}$ with dimension $\De$ in the IR CFT 
is to be viewed as order $E^\De$ in the low-energy expansion of dilaton
interactions in flat spacetime.
(CFT operators are defined to have definite dimensions,
and have vanishing VEV in flat spacetime.)
The most general relevant dilaton couplings to the CFT have the form
\beq[IRcoup]
\!\!\!\!\!\!\!
S_1[\scr{O}]
\eql{Ocoup}
&=\myint d^4 x \sqrt{-g} \,
(m \Om)^{4 - \De} \gap \scr{O}
\eeq
and
\beq
S_2[\scr{O}] &= \myint d^4 x \sqrt{-\hat{g}} \,
m^{2 - \De} R(\hat{g}) \gap \hat{\scr{O}}
\nonumber\\
\eql{ROcoup}
&=\myint d^4 x \sqrt{-g} \,
(m \Om)^{2 - \De} 
\left( R(g) - 6 \Om^{-1} \Box \Om \right) \! \scr{O},
\eeq
where $\scr{O}$ is a scalar primary operator with dimension $\De$.
In the low-energy expansion
\beq
S_1 = O(m^{4-\De} E^\De),
\qquad
S_2 = O(m^{2 - \De} E^{2 + \De}),
\eeq
so $S_1$ is relevant for $\De < 4$,
and $S_2$ is relevant for $\De < 2$.
It is easily seen that there are no other
relevant couplings.
If $\scr{O}$ is a non-primary operator then \Eq{Ocoup} is a
total derivative and \Eq{ROcoup} is irrelevant,
since unitarity bounds imply
\beq
\De(\Box \scr{O}) > 3,
\qquad
\De (\nabla_\mu \scr{J}^\mu) > 4,
\qquad
\De (\nabla_\mu \nabla_\nu \scr{T}^{\mu\nu}) > 6,
\eeq
{\it etc\/}.
We can write other dilaton couplings involving higher spin
operators such as
\beq
\De S = \myint d^4 x\, \sqrt{-\hat{g}} \left[
m^{1 - \De} \hat{\nabla}_\mu R(\hat{g}) \hat{\scr{J}}^\mu
+ m^{2 - \De} R_{\mu\nu}(\hat{g}) \hat{\scr{T}}^{\mu\nu} \right],
\eeq
but these are also irrelevant by the unitarity bounds.

Note that the IR CFT always has at least one
relevant operator, namely the identity
operator $\scr{O} = 1$.
The term $S_1[\scr{O} = 1]$ is a cosmological
constant term that gives contributions to dilaton-dilaton scattering
that are larger than the WZ term.
We eliminate this by tuning the IR cosmological constant to
zero.
The term $S_2[\scr{O} = 1]$ is a kinetic term
for the dilaton.
It is quadratic in $\Om$ (and hence $\varphi$),
and therefore does not contribute to dilaton scattering.

If the IR CFT contains relevant operators $\scr{O} \ne 1$,
$S_1[\scr{O}]$ must be fine-tuned
away, otherwise the CFT does not describe
the asymptotic behavior of the theory in the IR as assumed.
This leaves $S_2[\scr{O}]$, which gives a relevant coupling
of the dilaton to the CFT for $\De < 2$.
This cannot be fine-tuned away in general, but it does not
affect the dynamics of the CFT in flat spacetime, and
it does not contribute to dilaton scattering
if we impose the on-shell condition \Eq{onshell}.
Other possible effects of these terms will be discussed
in \S\ref{sec:Dele2} below.

Note that imposing the on-shell condition \Eq{onshell}
is equivalent to solving the equations of motion that
result from varying the kinetic term
$S_2[\scr{O} = 1]$.
We could therefore define the amplitude more physically by
giving a large coefficient $f^2$
to the kinetic term and expanding
in inverse powers of $1/f$.
This gives a nice physical interpretation to the dilaton
scattering amplitude, but is not strictly necessary for our argument.
We prefer to emphasize that nowhere in our arguments does the
dilaton need to be dynamical.

Finally, we must consider 4-derivative
terms that depend only on the dilaton.
At 4 derivatives, there are three independent Weyl invariant
terms that may be written
\beq[R2counterterms]
\De S = \myint d^4 x \sqrt{-\hat{g}} \left[
 E_4(\hat{g}) + W^2(\hat{g})+R^2(\hat{g}) 
\right].
\eeq
These however do not contribute to dilaton scattering in
flat spacetime:
the first term is a total derivative;
the second vanishes
identically on conformally flat metrics;
and the third has the form
\beq[R2counterterm]
\sqrt{-\hat{g}} \, R^2(\hat{g})
= \sqrt{-g} \left[ R(g) - 6 \Om^{-1} \Box \Om \right]^2,
\eeq
which does not contribute if we impose the on-shell condition
\Eq{onshell}.

We have therefore shown that the leading contribution to
the dilaton scattering amplitude comes from the WZ term \Eq{WZ}.

\subsection{Operators with $\De \le 2$
\label{sec:Dele2}}
We have seen above that if there is a scalar primary operator
$\scr{O}$ with dimension $\De < 2$ ($\De = 2$), the theory has a relevant (marginal) interaction
term of the form 
\beq[ROrelevant]
\De S 
= \myint d^4 x \sqrt{-\hat{g}} \,
m^{2 - \De} R(\hat{g}) \hat{\scr{O}}.
\eeq
These terms do not affect the dilaton-dilaton scattering amplitude
in flat spacetime on-shell, because $R(\hat{g}) = 0$ for these
backgrounds.
On the other hand, they do give IR singular contributions to the
correlation functions of $T^{\mu\nu}$. 
In this section we will discuss in more detail these operators 
and make it more clear that they do not invalidate our arguments.

If $\hat\scr{O}$ is a singlet under all symmetries, then the theory
is highly unnatural as an IR CFT.
However, there is one important special case where a theory
with such an operator appears in the IR, namely theories
with Nambu-Goldstone bosons.
These are free massless scalars $\pi$ in the IR, and the absence of a 
mass for the scalars is rendered natural by a $\pi$ shift symmetry.
This shift symmetry also forbids an improvement term of the form
$\pi^2 R(g)$ that makes the stress-energy tensor conformal.
We can write this theory as a conformal scalar plus an ``unimprovement''
term $-\frac 1{12} \pi^2 R(\hat{g})$,
where $\pi^2$ is a primary operator of dimension $\De = 2$.

To understand the effects of the term \Eq{ROrelevant} on the
off-shell dilaton amplitudes, note that it contains at least one power
of the dilaton field $\vph$.
This means that the $n$-dilaton amplitude involves at most $n$
insertions of the term \Eq{ROrelevant}.
If $\De < 2$, the term with the most insertions of the interaction
is the most relevant, so \eg\ the dilaton-dilaton scattering 
amplitude is
\beq
A(p_1, \ldots, p_4) \sim \left( \frac{m^{2 - \De}}{f} \right)^4
p_1^2 \cdots p_4^2 \,
\avg{\scr{O}(p_1) \cdots \scr{O}(p_4)}.
\eeq
The correlation function of $\scr{O}$ is to be evaluated in the
unperturbed CFT.
We see that the amplitude is singular in the IR, but vanishes on-shell, 
as it must. 
We can extend this logic to correlation functions of the full
stress-energy tensor $T^{\mu\nu}$.
The term \Eq{ROrelevant} vanishes in flat spacetime, so the
$n$-point functions of $T^{\mu\nu}$ in the perturbed theory
are given by a finite sum of $2, \ldots, n$ point functions of
the unperturbed theory.
This clarifies the relationship between the anomaly in the
perturbed and the unperturbed theory, namely the anomalous terms in the
perturbed theory are determined by the anomaly of the unperturbed theory
in a straightforward way.

It should be stressed that the extra terms discussed above correspond to
genuinely nonlocal effects in the quantum effective action $W[\hat g_{\mu\nu}]$.
This is not surprising since \Eq{ROrelevant} is an explict 
IR breaking of conformal invariance. 
We can see the nonlocal structure by direct
calculation of the quantum effective action in some simple cases.
For example,
in the simplest case where $ \hat{\scr{O}}= \Phi$ is a free scalar
field with $\De = 1$ the generated term is
\beq
\De W \sim \myint d^4 x\, \sqrt{-\hat g} \, m^2 R(\hat g)
\frac{1}{\hat\Box} R(\hat g),
\eeq
while in the case $\hat{\scr{O}}= \pi^2$ where $\pi$ is a free massless
Goldstone boson we get a series of terms 
\beq
\De W \sim \myint d^4 x\, \sqrt{-\hat g}\,
R(\hat g)(\ln \hat\Box )R(\hat g) +\cdots
\eeq
The Weyl variation of these term is also  non-local. That is not a contradiction since \Eq{ROrelevant} 
explicitly breaks Weyl invariance. However an on-shell dilaton is beautifully insenstive to this effect.

We also remark that in the presence of the term \Eq{ROrelevant}, 
the stress-energy tensor defined by variation with respect to the metric is 
not the canonical  energy momentum tensor of the IR CFT 
(which is a primary spin 2 field). 
It differs from it by mixing with descendants of $\hat{\scr{O}}$.
Roughly one has
\beq[TUVtoTIR]
T^{\mu\nu} = T^{\mu\nu}_{\rm IR\,CFT}+ m^{2-\De} 
(\d^\mu \d^\nu - \Box \eta^{\mu\nu}) {\cal O} + \cdots \, .
\eeq
However, all of these effects do not matter  
in backgrounds with $\Box\Om = 0$.

Finally, we show that even if operators with $\De \le 2$
can be somehow tuned away in the action,
they give rise to IR divergences in the off-shell dilaton-dilaton scattering 
amplitude.
We have
\beq
A(p_1, \ldots, p_4) &= 
\frac{\de^4 W}{\de \vph(p_1) \cdots \de \vph(p_4)}
\nonumber\\
&= \avg{T(p_1) T(p_2) T(p_3) T(p_4)}
+ \mbox{contact terms}.
\eeq
For the forward amplitude $p_1 = -p_3$, $p_2 = -p_4$, 
the potential divergence is in the zero momentum channel, 
corresponding to $x_1 - x_3 \sim x_2 - x_4 \ll x_3 - x_4$.   Let us analyze this first in the regime $x_1 - x_3 \sim x_2 - x_4 \gg m^{-1}$.  In the effective IR CFT the trace has dimension $\De_{\rm IR} > 4$, so  
 the OPE for the two close pairs gives
\beq
A &\propto m^{16 - 4\De_{\rm IR}} \myint d^4 x_1 \, d^4 x_2 \, d^4 x_3 \, 
e^{i [p_1 \cdot (x_1 - x_3) + p_2 \cdot (x_2 - x_4)]} 
\nonumber\\
&\qquad\qquad\qquad
{} \times (x_1 - x_3)^{\De - 2 \De_{\rm IR}} (x_2 - x_4)^{\De - 2 \De_{\rm IR}}
\langle {\cal O}(x_3) {\cal O}(x_4) \rangle
\nonumber\\
&\propto m^{16 - 4\De_{\rm IR}} (m^{2 \De_{\rm IR} - \De - 4} - p_1^{2 \De_{\rm IR} - \De - 4} )(m^{2 \De_{\rm IR} - \De - 4} - p_2^{2 \De_{\rm IR} - \De - 4})\nonumber\\
&\qquad\qquad\qquad
\times \int d^4x_3\, (x_3 - x_4)^{-2\De_{\cal O} } \,,
 \eeq
where $\cal O$ is the operator of lowest dimension $\De$ in $TT$.  
In each parenthesis, the term $m^{2 \De_{\rm IR} - \De - 4}$ comes from the UV end of the integral.  This term cannot actually be present it corresponds to a effective interaction proportional to $m^{4 - \De} {\cal O}$ in the IR CFT coupled to a background metric.  Since the IR is a CFT this must be absent, canceled by the tuning that makes the IR theory conformal.  The remaining contribution is proportional to positive powers of $p_1^2$ and $p_2^2$: we see that if  $\De \leq 2$ the off-shell amplitude diverges, but the on-shell amplitude is IR finite.  
(The apparent divergence at $p_i^2 = 0$ for $\De > 4$ is not real: 
it comes from $x_1 - x_3\,, x_2 - x_4 \gg x_3 - x_4$ where the OPE 
is not valid.)  The regime $x_1 - x_3 \sim x_2 - x_4 \lsim m^{-1}$ may similarly bring in powers of the UV cutoff, which must be absent by renormalization.

\subsection{Dispersive Argument
\label{sec:disperse}}
We now consider dilaton scattering at low energy.
Specifically, we are interested in the amplitude
for $\varphi\varphi \to \varphi\varphi$ as a function of
the Mandelstam variable $s$, with $t \to 0$.
We denote this by $A(s)$.
To define the amplitude, it is convenient to use our
freedom to add UV counterterms to the theory that depend
only on $\hat{g}_{\mu\nu}$.
We will cancel the cosmological constant in the IR by
adding a suitable counterterm, and we will
add a WZ term to cancel the WZ term
induced from the UV theory.
That is, we compute the dilaton scattering amplitude using
\beq
W[\hat{g}_{\mu\nu}]
+ S_{\rm WZ}[g_{\mu\nu}, \tau; a_{\rm UV}, c_{\rm UV}].
\eeq
This gives an amplitude with good behavior 
in the UV and IR, allowing the use of unsubtracted dispersion
relations to establish the positivity of the amplitude.
Alternatively, we could use a subtracted dispersion relation
on the amplitude defined by $W[\hat{g}_{\mu\nu}]$.

The leading behavior as $s \to 0$ of the amplitude is then
\beq[Asmalls]
A(s) \to \al \frac{s^2}{f^4} + O\left( \frac{m^{2(4 - \De_{\rm IR})} 
s^{\De_{\rm IR} - 2}}{f^4} \right),
\eeq
where second term arises from the coupling of the dilaton to the
IR CFT.
Here $\De_{\rm IR} > 4$ is the lowest dimension of the irrelevant
operators appearing in the deformation of the IR CFT.
Note that if we had not canceled the cosmological constant term
in the IR, the leading behavior of the amplitude would be $\sim s^0$.

As $s \to \infty$ the UV behavior is dominated either by the
largest dimension relevant deformation with dimension $\De_{\rm UV}$,
or by the cosmological constant term:
\beq[Alarges]
A(s) \sim 
\begin{cases}
\displaystyle \frac{m^{2(4 - \De_{\rm UV})} 
s^{\De_{\rm UV} - 2}}{f^4}
& if $\De_{\rm UV} \ge 2$, \cr
\displaystyle \frac{m^4}{f^4} 
& if $\De_{\rm UV} < 2$.
\end{cases}
\eeq
Note that if we had not subtracted the UV WZ term,
the leading behavior of the amplitude would be $\sim s^2$.

With this definition of the amplitude, 
the coefficient of the WZ term in the IR is
$a_{\rm UV} - a_{\rm IR}$, so we have
\beq
\al = 8( a_{\rm UV} - a_{\rm IR} ).
\eeq
Therefore, the $a$-theorem is equivalent to $\al > 0$.

Theories with $\al < 0$ have superluminal propagation of dilaton
excitations in certain nontrivial background dilaton configurations,
and are therefore acausal.
However, in order for this to be a physical problem,
the dilaton must be a propagating degree
of freedom.
The acausality of this theory may simply be a sign that we
cannot UV complete the theory with a dynamical dilaton.%
\footnote{In fact, note that we can get any value for
$\al$ by choosing the WZ counterterm in the UV arbitrarily.
The choices that give $\al < 0$ presumably do not have
causal UV completions.}
In fact, we do not know of any non-supersymmetric UV
completion of a theory with a dynamical dilaton.

We will give a rigorous dispersive
argument that $\al > 0$ without assuming
anything about the UV completion of the theory.
We consider the contour integral
\beq[semicon]
0 = \frac{1}{2\pi i} \oint ds \, \frac{A(s)}{s ^3} 
\eeq
along the contour shown in Fig.~2.
\begin{figure}[t]
\begin{center}
\includegraphics[scale=1]{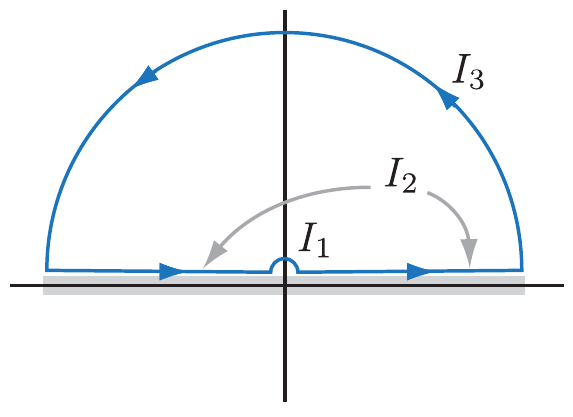} 
\end{center}
\caption{The integration contour in the complex $s$ plane used
to argue that $\al > 0$.}
\end{figure} 
The integral $I_1$
over a small semi-circle of radius $\ep$ is given by
\beq[I1]
I_1 = -\frac{\alpha}{2f^4} + O(\ep^{\De_{\rm IR} - 4}).
\eeq
Because $\De_{\rm IR} > 4$, the $\ep \to 0$ limit
picks out the coefficient
of the leading low-energy behavior of the scattering amplitude.

We now turn to $I_2$.
The function $A(s)$ has cuts all along the real $s$ axis due to
massless CFT intermediate states.
By crossing symmetry $A(s) = A(u) = A(-s)$,
and so the contribution from the integral along the real axis is
\beq
I_2 = \frac{1}{\pi} \int_\ep^\infty ds \,
\frac{\Im A(s)}{s^3}
= \frac{1}{\pi} \int_\ep^\infty ds\,
\frac{ \si(s)}{s^2} \,,
\eeq
where $\si$ is the total cross section for $\varphi\varphi \to \mbox{CFT}$
in the probe limit.
The integral $I_2$ is manifestly positive, so we have proved $\al > 0$
(and hence the $a$-theorem) provided that we can neglect the contribution
from the large semicircle.
In fact, the UV behavior \Eq{Alarges} is precisely sufficient
to ensure that this is the case.

This completes the proof of the $a$-theorem in the case where the
UV CFT is deformed by relevant operators, and the IR CFT by irrelevant
operators.
In the remainder of this section, we consider
the important special case where the flow in the UV or
IR is logarithmic, induced by marginally relevant or irrelevant operators.
For example, we may be interested in an asympotically free gauge theory
in the UV, or theories with $U(1)$ gauge factors in the IR.
We will show very generally that
the dilaton decouples sufficiently rapidly that we can
apply the dispersive argument above to these cases.
The issue amounts to the UV and IR convergence of the integral
$I_2$, since the IR convergence implies that $I_1$ is
given by \Eq{I1}, and the UV convergence implies $I_3 = 0$.

We consider first the case where the UV theory is
an asymptotically free gauge theory.
This theory requires a UV cutoff $\La$,
which breaks scale (and conformal) invariance and induces a coupling
to the dilaton.
The dependence on $\La$ is 
governed by the RG equation
\beq[rge]
\frac{d}{d\ln\La} \left( \frac{1}{g^2} \right) = b + O(g^2)
\eeq
with $b > 0$.
We write the UV Lagrangian as
\beq
{\scr L}_{\rm UV} = 
-\frac{1}{4g^2}
F_{\mu\nu}^2
+ \mbox{regulator\ terms}.
\eeq
where $F_{\mu\nu}$ has no $g$ dependence.
We can couple the dilaton to make the theory scale invariant
by making the replacement 
\beq
\La \to \La \Om
\eeq
in the regulator terms.
The dependence on the dilaton then follows from the
the RG:
\beq
{\scr L}_{\rm UV} &\to {\scr L}_{\rm UV} + 
\left[ \frac{d}{d\ln\La} \left( \frac{1}{g^2} \right) \ln \Om
+ \frac 12 \frac{d}{d(\ln\La)^2} \left( \frac{1}{g^2} \right) \ln^2 \Om
+ \cdots \right] F_{\mu\nu}^2
\nonumber\\
&= \scr{L}_{\rm UV} +
\left[  -\frac{b}{4} \ln \Om + O(g^2) \right]
F_{\mu\nu}^2.
\eeq
When expanded in $\varphi$, this coupling
contains all powers of the dilaton excitations.
The quantity of interest is the
cross section for two dilatons to scatter into two 
gauge bosons.
We canonically normalize the gauge fields by
$A_\mu \to g A_\mu$ and find the amplitude
$\scr{M}(\varphi\varphi \to AA) \sim b g^2 / f^2$.
For $s \to \infty$ we therefore have
\beq[dildilAA]
\sigma (\varphi\varphi \to AA) \sim \frac{b^2 g^4(s)}{f^4} s\,.
\eeq
The solution of the RG equation \Eq{rge} is
$g^2(s) \sim 1/\ln s$,
so the integral $I_2$ has the large $s$ behavior 
\beq
I_2 \sim 
\int^\infty \frac{d\ln s}{(\ln s)^2}
\eeq
which converges in the UV.
For $I_3$ we need the large-$s$ behavior of the real as well
as the imaginary part of $A$.
In order to get the correct imaginary part we must have
$\Re(A) \sim \frac{s^2}{\ln s}$, which is just sufficient to
give $I_3 = 0$.

A similar argument holds if the IR theory has a gauge
coupling that runs to zero in the IR.
In that case, $g^2(s) \sim 1/\ln s^{-1}$ as $s \to 0$, and we have
\beq
{\rm Im}\,A(s) - \frac{\al s^2}{f^4} \sim
\frac{b^2 g^4(s)}{f^4} s^2 
\sim \frac{s^2}{(\ln s^{-1})^2}.
\eeq
The integral therefore has the small $s$ behavior
\beq
I_2 \sim \int_0 \frac{d \ln s^{-1}}{(\ln s^{-1})^2},
\eeq
which is sufficient for convergence in the \mbox{IR}.
If the one-loop beta function vanishes, we have
\beq
\frac{d}{d\ln \La} \left( \frac{1}{g^2} \right)
= b' g^{2}
\eeq
with $b' > 0$.
In this case, the amplitude
$\scr{M}(\varphi\varphi \to AA) \sim b' g^4 / f^2$
and we have as $s \to 0$
\beq
A(s) - \frac{\al s^2}{f^4}
\sim \frac{b'^2 g^8(s)}{f^4} s^2
\sim \frac{s^2}{(\ln s^{-1})^2},
\eeq
as before.

Gauge theory is an important special
case, but it is somewhat unusual from the CFT point of view
because the dimension 4 operator that generates the flow
is $F_{\mu\nu}^2$,
and the free limit corresponds to an \emph{infinite}
coefficient for this operator in the action.
We therefore consider the effect of adding a dimension 4
operator $\scr{O}$ with coefficient $\la$ to either the UV or the
IR CFT:
\beq
\De\scr{L} = \la \scr{O}.
\eeq
For small $\la$ the 
RG equation has the form
\beq
\frac{d\la}{d\ln\La} = b \la^n.
\eeq
For example for a $\Phi^4$ term $n = 2$, while for a Yukawa
coupling $n = 3$.
The dilaton coupling follows from the replacement
$\La \to \La(1-\varphi/f)$, so we have
$A(\varphi\varphi \to \mbox{QFT}) \sim b \la^n/f^2$ and hence
\beq
\si(\varphi\varphi \to \mbox{QFT}) \sim \frac{b^2 \la^{2n}(s)}{f^4} s.
\eeq
For the case where $b > 0$ and this operator appears in the
IR CFT, we have as $s \to 0$
\beq
\la(s) \sim \left( \frac{1}{\ln s^{-1}} \right)^{1/(n-1)}
\eeq
and the integral $I_2$ has the form as $s \to 0$
\beq
I _2\sim \myint \frac{d\ln s^{-1}}{(\ln s^{-1})^{2n / (n - 1)}}.
\eeq
This always converges as $s \to 0$ because $2n / (n - 1) > 2$
for $n > 1$.
In the case where the operator
$\scr{O}$ appears in the UV CFT, a similar argument
proves the UV convergence. 
Similarly one can see that $|A(s)/s^2|\to 0$  at infinity, 
ensuring $I_3=0$.
We conclude that the convergence of the dispersion relation
holds very generally even with logarithmic flows in the
UV or IR.

\subsection{Gravitational Anomalies}

We have assumed that it is possible to couple to a background metric in a 
diffeomorphism invariant way.
This is not possible if there is $U(1)$ gauge symmetry with
${\rm Tr}\, Q \neq 0$, where $Q$ is the gauge charge.
In such theories, the gauge current is not conserved
if we require diffeomorphism invariance:
\beq
\d_\mu J^\mu = -\frac{\tr(Q)}{384\pi^2} R \tilde{R},
\eeq
where
$ R \tilde{R} = \frac 12 \ep^{\mu\nu\rho\si} R_{\mu\nu}{}^{\tau\om}
R_{\rho\si\tau\om}$.
Alternatively, we can maintain gauge invariance at the
price of diffeomorphism invariance by using the fact that
\beq
R \tilde R = \d_\mu K^\mu
\eeq
where $K^\mu$ is not generally covariant.
We add to the diffeomorphism invariant action a term
\beq[gaugecorrection]
\De S = \frac{\tr(Q)}{384\pi^2} 
\myint d^4 x\, \sqrt{-g}\, A_\mu K^\mu.
\eeq
The fact that this term violates diffeomorphism invariance
is not a fundamental problem for us, since the metric is just
a background field.
It does mean that we must take care in defining 
the background metric, since it no longer has a geometrical
meaning.
For our purposes, it is sufficient to write
\beq
g_{\mu\nu} = e^{-2\tau} \eta_{\mu\nu}
\eeq
where $\eta_{\mu\nu}$ is the flat metric in Cartesian 
coordinates and $\tau$ is the dilaton field.

Now the point is that the
additional term \Eq{gaugecorrection} does not contain any 
coupling of the dilaton.
The reason is that for the background metric $e^{-2\tau} \eta_{\mu\nu}$
this coupling must respect global Lorentz invariance.
It is a polynomial involving $\ep^{\mu\nu\rho\si}$,
$A_\mu$, 3 derivatives, and powers of $\vph$.
It is easy to see that there is no such Lorentz invariant, 
and so the dilaton coupling vanishes identically.
The presence of this term therefore does not affect the arguments
of this paper, very much like the improvement terms for scalars,
and the $R^2$ counterterms.

Alternatively, we can avoid breaking diffeomorphism invariance 
by introducing a background 2-form field field $B_{\mu\nu}$,
and restore diffeomorphism invariance by the Green-Schwarz 
mechanism~\cite{Green:1984sg}.  
The anomalous variation is of the form 
\beq
\de W \propto \int \! F \!\wedge\! {\rm Tr}(\de \lambda R) ,
\eeq
where $\de\lambda$ is a local Lorentz transformation.
Introduce an interaction $\int\! F \!\wedge\! B$.  
This is invariant under the usual 2-form transformation 
$\de B = d\xi$, and with the transformation 
$\de B = -  {\rm Tr}(\de \lambda R)$ it cancels the anomaly.  
The complication is that one must now consider possible gauge-invariant 
interactions built out of the field strength 
$H = dB + \omega_{3\rm L}$, where $\omega_{3\rm L}$ is the gravitational 
Chern-Simons  term.
In fact, $H$ vanishes in the background $B=0$, $\hat g = \eta e^{-2\tau}$, 
and so the extra terms do not contribute to dilaton amplitudes.

\section{Restrictions on Perturbative RG Flows
\label{sec:pertRG}}

We now generalize these arguments to restrict perturbative RG flows.
We will show that in unitary theories the only IR or UV asymptotics
that can be described in pertrubation theory is conformal invariance.
Our arguments in fact apply to small perturbations of a conformal
fixed point, even if it strongly coupled.
Closely related monotonicity results for perturbative flows were 
obtained in \Refs{Osborn:1989td,Jack:1990eb,Osborn:1991gm}.

\subsection{General Framework
\label{sec:framework}}

Our argument is based on the same amplitude $A(s)$ defined above.
We work with the dimensionless function $\al(s)$ defined by
\beq
A(s) = \frac{\al(s) s^2}{f^4}.
\eeq
It will be essential to our argument below that this amplitude
has no counterterm, and is therefore calculable purely in terms
of the renormalized couplings in the theory; this follows from the analysis in \S2 of possible dilaton couplings.  To reiterate the basic idea, we have included the most general 
diffeomorphism invariant counterterms depending on the background
metric $\hat{g}_{\mu\nu}$, and these do not affect the dilaton amplitude.
The split into $g_{\mu\nu}$ and $\tau$ used to define the dilaton
amplitude is arbitrary; the theory only depends on $\hat{g}_{\mu\nu}$.
The WZ term parameterizes the effects of the anomaly and does not
represent a counterterm.

In free field theory the function  $\al(s)$ equals $-8$ times the 
(constant) coefficient of the WZ term. 
The discussion of \S\ref{sec:dilatoncouplings} showed that there is no 
counterterm for the amplitude $A(s)$, and therefore all divergences in
$\al(s)$ can be absorbed into renormalization of the renormalized
couplings of the theory.
When interactions are turned on, $\al(s)$ will thus
be written as a power series in the renormalized couplings evaluated
at a renormalization scale $\mu \sim s^{1/2}$.
The same structure holds for marginal perturbations around any
conformal fixed point, with the couplings replaced by the coefficients
of the marginal operators in the perturbation.

In this section we are considering flows that remain perturbative 
in the UV or IR limit, that is, 
the coupling remains in some bounded neighborhood of the origin.  
The magnitude of $\alpha(s)$ is then bounded.  
This is the key observation that allows us to obtain new constraints 
on  RG flows.

On the closed contour $C$ of Fig.\ 2 we have
\beq
\frac{1}{i\pi} \int_C \frac{ds}{s} \,\al(s) = 0 \,.
\eeq
Let the radii of the inner and outer semicircles be $\si_1$ and $\si_3$,
respectively.
Then
\beq[albar]
\bar\alpha(\sigma_3) - \bar\alpha(\sigma_1) = - \frac{2}{\pi} \int_{\sigma_1}^{\sigma_3} \frac{ds}{s} \Im\al(s) \,,
\eeq
where $\bar\alpha(\sigma)$ is the average value of $\alpha(s)$ on a semi-circular contour of radius $\sigma$.
We have used $\al(-s) = \al(s)$,
which implies $\Im\al(-s') = -\Im\al(s')$.  We can also write this in a differential version
\beq[aldiff]
\sigma \partial_\sigma \bar\alpha(\sigma) = - \frac{2}{\pi} \Im\al(\sigma) \,,
\eeq

Now consider \Eq{albar} as $\sigma_1 \to 0$ with $\sigma_3$ fixed.  
The quantity $\bar\al(\si_1)$ is calculable in perturbation theory,
and must therefore remain bounded as long as perturbation theory is valid.
Therefore, the integral 
\beq[IRUVint]
\al(s) = \frac{2}{\pi} \int_{\sigma_1}^{\sigma_3} ds'\, \frac{\Im\al(s')}{s'}
+ \mbox{finite},
\eeq
must remain bounded as $\sigma_1 \to 0$.
Similarly, by taking $\sigma_3 \to \infty$ at fixed $\sigma_1$, we can conclude that the integral remains bounded as $\sigma_3 \to \infty$. 
The boundedness of these integrals has strong implications
in unitary theories, because for
these $\Im\al(s)$ is a sum of positive terms,
each of which is a squared amplitude connecting the
vacuum with some $n$-particle state.  
In particular it follows immediately that $\Im\al(s)$ vanishes in both 
the UV and IR limits.%
\footnote{In a non-unitary theory, there would be the possibility that that 
the integral \Eq{IRUVint} converges due to rapid oscillations
even if $\Im\al(s)$ is nonzero in
the IR or UV.}

In detail, we write the (in general off-shell)
dilaton-dilaton scattering amplitude
\beq
(2\pi)^4 \de^4(p_1 + \cdots + p_4) A(p_1, \ldots, p_4) =
\frac{\de^4 W}{\de \vph(p_1) \cdots \de \vph(p_4)},
\eeq
where
\beq[ATTTT]
f^4 A(p_1, \ldots, p_4) &= \avg{T(p_1) T(p_2) T(p_3) T(p_4)}
\nonumber\\
& \qquad
{}+ \avg{T(p_1 + p_2) T(p_3) T(p_4)}
+ \mbox{permutations}
\nonumber\\
& \qquad
{}+ \avg{T(p_1 + p_2) T(p_3 + p_4)}
+ \mbox{permutations}
\nonumber\\
\eql{AintermsofT}
& \qquad
{}+ \avg{T(p_1 + p_2 + p_3) T(p_4)} + \mbox{permutations} \,.
\eeq
(The average here means the time-ordered vacuum expectation value).
We have defined the trace of the stress-energy tensor by 
differentiating the quantum effective action with respect to $\tau$:
\beq[goodT]
\!\!\!\!\!\!\!
\avg{T(x_1) \cdots T(x_n)}
= \frac{1}{\sqrt{-g(x_1)}} \cdots
\frac{1}{\sqrt{-g(x_n)}} \,
\frac{\de^n W[e^{-2\tau} g_{\mu\nu}]}
{\de \tau(x_1) \cdots \de \tau(x_n)} \,.
\eeq
This definition coincides with the standard one,
\beq[standardT]
\!\!\!\!\!\!\!
\avg{T(x_1) \cdots T(x_n)}
= \frac{g^{\mu_1\nu_1}(x_1)}{\sqrt{-g(x_1)}}
\cdots \frac{g^{\mu_n\nu_n}(x_n)}{\sqrt{-g(x_n)}}\,
\frac{\de^n W[ g_{\mu\nu}]}
{\de g^{\mu_1\nu_1}(x_1) \cdots \de g^{\mu_n\nu_n}(x_n)}\,,
\eeq
up to contact terms.
Several different powers of $T$ are present in the amplitude~(\ref{eq:ATTTT}) because of the nonlinear relation~(\ref{eq:dildef}) between $\varphi$ and $\tau$.

The imaginary part of $\al(s)$ is therefore
\beq[Imalphastatesum]
\Im\al(s) = \sum_{X} 
\left| \bra{X}\, T(p_1) T(p_2) + T(p_1 + p_2) \,\ket{0} \right|^2 \,,
\eeq
summed over all states of the CFT.  
We are considering theories of massless particles, but there
are no IR divergences in
\Eq{Imalphastatesum} because these involve sums over all final states, 
\ie\ it is completely inclusive. 

To get constraints on the RG flow, we relate $\Im{\al}$ to the beta functions.
For any perturbative theory, we have the operator identity 
\beq[TequalsbetaO]
T = \sum_A B_A \scr{O}^A + \scr{E} + \mbox{anomaly terms} \,.
\eeq
The meaning of the various terms on the \rhs is as follows.
The operators $\scr{O}^A$ are a basis for the interaction terms in the
Lagrangian
\beq
\scr{L}_{\rm int} = \sum_A \la_A \scr{O}^A \,,
\eeq
and $B_A$ is a beta function associated with the coupling $\la_A$.  We are following here the notation of  \Refs{Osborn:1989td,Jack:1990eb,Osborn:1991gm}, which uses $B_A$ for the coefficients in $T$ when expressed in an operator basis where the divergence of currents, that in general appear on the right hand side of \Eq{TequalsbetaO}, are replaced by using the corresponding Ward identities. That  determines a shift from the``naive'' beta function
 $\be_A$ to $B_A$ in the coefficients of $\scr{O}_A$. As already noticed in \Ref{Osborn:1991gm}\footnote{See for instance the discussion in the introduction and below  Eq.~(4.31) in that Ref..}, and as explained in \S3.4 below, $\be_A$ have a degree of arbitrariness, which arises ultimately because the RG equation only involves 
$\int d^4x\, T$.
This subtlety will not affect our argument, which only uses
\Eq{TequalsbetaO} and therefore only refers to $B_A$.\footnote{In version 1 of the present work, the notation $\be_A$ was used for the coefficients in $T$.  We have changed this to conform with the notation of \Refs{Osborn:1989td,Jack:1990eb,Osborn:1991gm}.}

The term $\scr{E}$ in \Eq{TequalsbetaO} represents terms that
vanish by the equations of motion.
These are important for contact terms, \eg
\beq[contact]
\avg{\scr{E}(x) \scr{O}(y)} \propto \de^4(x - y) \avg{\de\scr{O}(x)},
\eeq
where $\de\scr{O}$ is the
infinitesmal variation of $\scr{O}$ under scale transformations.
The anomaly terms in \Eq{TequalsbetaO} are proportional to
\beq
a E_4(g) - c W^2(g),
\eeq
and are nonzero only for a nonzero gravitational field.

The operator identity \Eq{TequalsbetaO} is in general valid for
single insertions of $T$ in correlation functions involving arbitrarily
many powers of elementary fields.
When $T$ is inserted more than once, there are additional contact
terms that can be thought of as additional insertions of scalar
operators.  
In \Eq{Imalphastatesum} these are important because of the
appearance of the operator product $T(p_1) T(p_2)$.
In \S3.2 we avoid this complication by focusing on intermediate states 
$\ket{X}$ with nonzero angular momentum.  
In \S\ref{sec:j=0} to \S\ref{sec:wzcon}
we include also the effects of the $J = 0$ states, 
giving somewhat sharper constraints. 

We also note that the discussion in this section
and the following ones can be directly generalized to
perturbation theory around any conformal fixed point, 
whether free or strongly interacting.
For perturbations around an IR fixed point,
the $\De > 4$ operators flow to zero
and the $\De < 4$ operators must be tuned away in order for the
perturbative to describe the IR limit
(like scalar mass terms in in the free conformal theory).
The only nontrivial effects therefore come from $\De = 4$ operators.%
\footnote{This can be generalized to operators where
$\De - 4$ is less than or of order the dimensionless quantity
describing the perturbation, as is the case in Wilson-Fisher fixed
points.}
For perturbations around a UV fixed point, 
we must require that the $\De > 4$ operators are
not present, while the $\De < 4$ operators flow to zero in the IR,
so again the only nontrivial effects come from dimension-4 operators.
For such theories, we expect \Eq{TequalsbetaO} to
hold, where $\scr{O}^A$ are a complete set of dimension-4 primary
operators.
Further details will be presented in \Ref{BoazR}.
 
\subsection{Constraints from Higher Partial Waves
\label{sec:higher}}
Restricting to $J > 0$ partial waves eliminates both the contact interactions~(\ref{eq:contact}) and the linear $T$ term~(\ref{eq:Imalphastatesum}), 
leaving
\beq
\!\!\!\!\!\!\!
\Im\al(s) &\ge  \sum_{X, J_X > 0}
\bigl|  \bra{X} T(p_1) T(p_2) \ket{0} \bigr|^2
\\
&= \sum_{A,B} \left[ B_A(s) B_B(s) \right]^2
 \sum_{X, J_X > 0}
\bigl|  \bra{X} \scr{O}^A(p_1) \scr{O}^B(p_2) \ket{0} \bigr|^2
\eeq
This is a sum of positive terms, so each term must 
go to zero as $s \to 0$ and as $s \to \infty$.
For example, we can consider only the $A = B$ terms and conclude
that $B_A \to 0$ for all $A$ as $s \to 0$ provided that
\beq[nJterm]
\bigl|  \bra{X} \scr{O}^A(p_1) \scr{O}^A(p_2) \ket{0} \bigr|^2
\ne 0
\eeq
as $s \to 0$ for some $X$ with $J \ne 0$.
But this can easily seen to be the case by considering an
intermediate  state for which there is a tree-level matrix
element with $\scr{O}^A \scr{O}^A$.
For example, if $\scr{O}^A$ is a $\Phi^4$ interaction, we consider a tree level graph where the two operators are connected by a single propagator and the other legs connect to a six-scalar intermediate state: this contributes to all partial waves.  Similarly, for $\scr{O}^A$ a Yukawa interaction a graph with a scalar exchange contribute for four-fermion intermediate states.
Recalling that $\al(s)$ and the beta functions $B_A$ are all
dimensionless, the sum over $X$ (implicitly including phase space integrals) will give a
nonzero dimensionless constant for such states.

We conclude that $B_A$ must vanish for all $A$ as $s \to 0, \infty$, sufficiently rapidly that 
\beq[b4conv]
\int \frac{ds}{s} B_A^2 B_B^2
\eeq
converges.
Recall again that we are assuming that perturbation theory is valid
in the asymptotic IR.
Of course, there are theories (like QCD) where the couplings
get large in the IR, invalidating the use of perturbation theory
in the IR limit.
We cannot draw any conclusions about such theories.
But for theories that remain perturbative in the IR limit,
we have shown that $T \to 0$ as an
operator statement as $s \to 0$.
By definition, this means that the theory is conformal
in the IR.

Similar arguments can be made for the UV limit.
These show that  the UV limit is conformal if it is 
described in perturbation theory.
Note that an implicit assumption of this argument is
that the asymptotics is governed by a massless theory
with a fixed Lagrangian, which does not allow theories
with infinite numbers of massive particles coming in
at higher and higher scales.
Such theories are of course perfectly physical, 
so our constraint here is less general.

\subsection{Constraints from $J = 0$
\label{sec:j=0}}

Now let us consider the full sum over states~\Eq{Imalphastatesum}, 
focusing in particular on the $J=0$ partial wave previously omitted.  
We must then consider contributions from the matrix element of
$T(p_1 + p_1)$ as well as $T(p_1) T(p_1)$.
In perturbation theory, we expect the $T(p_1 + p_1)$ terms to dominate because each
derivative with respect to $\tau$ is proportional to a beta function.
This is indeed correct, but the argument is somewhat subtle because we must exclude
the possibility that the contact term in the $T(p_1) T(p_2)$ term has a
contribution that is linear in beta functions.
In \S\ref{sec:dimreg} we will exclude this possibility using dimensional 
regularization, and in \S\ref{sec:wzcon} we will derive an equivalent 
result from Wess-Zumino consistency conditions.%

We will proceed assuming that the $T(p_1 + p_2)$ term dominates in \Eq{Imalphastatesum}.
This term can be written
\beq[AT]
s^2 \Im\alpha(s) &= \Im \avg{T(p) T(-p)} \nonumber\\
&=  \sum_{A,B} B_A(s) B_B(s) 
\Im \avg{
 \scr{O}^A(p) \scr{O}^B(-p)} 
\eeq
where $p^2 = s$.
For a perturbative theory the two-point function 
can be approximated by free field theory,
\beq
 \avg{
 \scr{O}^A(p) \scr{O}^B(-p)}
 = i c_A \de^{AB} s^2 \ln\left( -\frac{s+i\epsilon}{\mu^2} \right) \,.
 \eeq
 \Eq{aldiff} then becomes
\beq[alrun]
r \partial_r \bar\alpha(r) = -2 \sum_A c_A B_A^2 \,,
\eeq
up to terms higher order in the beta functions,
where $c_A$ are coefficients that are positive by unitarity.
Specifically, for 
\beq
\bal
\scr{O}^1 &= \frac{1}{4!} \Phi^4 \,,
\qquad\ \ \ c_1 =  \frac{1}{{2^{10} (4!)^2 \pi^6}} \,, \\
\scr{O}^2 &=  \Phi \bar\Psi \Psi \,,
\qquad\ \ \ c_2 =  \frac{1}{2^4 4! \pi^4} \,, \\
\scr{O}^3 &= F_{\mu\nu}^2 /4g^4\,,
\qquad c_3 =   \frac{1}{2^5\pi^2 g^4} \,,
\eal
\eeq
where $\Psi$ is a Dirac fermion.
This implies that the flow of $\bar \al{\si}$ is monotonic.
Because $\bar\al$ coincides with $a$ at conformal fixed points,
this immediately gives a perturbative proof of the $a$-theorem.%
\footnote{In version 1 of this paper, \Eq{alrun}
was interpreted as an RG equation for the coefficient $a$ of the WZ term.  
It was pointed out in Ref.~\cite{FGSJO} that this interpretation is not correct.
However, the conclusions we draw from monotonicity of $\bar\al$ are not
affected by this.}
A very similar differential equation was obtained in 
\Refs{Osborn:1989td,Jack:1990eb,Osborn:1991gm}, as explained
below in \S3.5.

Using the same arguments given in \S\ref{sec:higher},
we can conclude that the integral
\beq[b2conv]
\int \frac{ds}{s} B_A^2 
\eeq
converges both in the UV and the IR, provided the couplings remain perturbative.
This conclusion is somewhat stronger than \Eq{b4conv} derived previously.

\subsection{B versus $\be$}
\label{sec:Toperator}
In this section we  explain the distinction between $\be_A$
and $B_A$ in \Eq{TequalsbetaO} and comment on the significance
of limit cycles as found in \Refs{FGSexample,FGSta}.
To illustrate the basic idea we focus on
general $\Phi^4$ theory with  $N$ flavors whose renormalized Lagrangian  
(in flat space) is
\beq
{\cal L}=\frac{1}{2}\partial_\mu \Phi_i\partial^\mu \Phi_i+\frac{\lambda_{ijk\ell}}{4!}\Phi_i\Phi_j\Phi_k\Phi_\ell\, .
\eeq
The $\be$-functions and the anomalous  dimensions of the fields are defined by expanding the trace of the canonical
energy momentum tensor $T_\mu^\mu$ in a complete basis of renormalized composite operators. 
In the above  theory, generalizing for instance
\Ref{Brown:1979pq},  and indicating by $[{\cal O}]$ the renormalized operator
one can write
\beq
T = {\be_{ijk\ell}}[{\cal O}_{ijk\ell}]
+ N_{ij}\partial_\mu [J^\mu_{ij}]
+ \Ga_{ij}[\phi_j E_i]+a_{ij}\Box[\phi_i\phi_j]
\eql{trace}
\eeq
where
${\cal O}_{ijk\ell}= \frac{1}{4!} \Phi_i\Phi_j\Phi_k\Phi_\ell$, 
 $J^\mu_{ij}$ is the flavor current,
 and $E_i=\de S/\de \Phi_i$ is the equation of motion operator. 
 The coefficient of the third term defines the dimension matrix: 
 $\Ga_{ij}=\de_{ij} +\gamma_{ij}$.
 Notice that the second term is absent in the case of the single flavor $\Phi^4$ theory of \Ref{Brown:1979pq}, but is generally expected by power counting: $N_{ij}$ is an antisymmetric $O(N)$ tensor  covariantly constructed from the coupling tensor $\lambda_{ijk\ell}$. 
 The last term can be improved away and plays no role in this discussion 
 and in dilaton amplitudes. 
Note that the response of the theory to a global scale
transformation is given by an insertion of
$\int d^4 x\, T(x)$,
in which the second and fourth terms in \Eq{trace} drop out
because they are total derivatives.
This gives the Callan-Symanzik equations for the theory.

The local \Eq{trace} is what matters to describe the effect of conformal transformations, \ie~local dilations. 
However there is an inherent ambiguity in the definition of $\be, N, \Gamma$.
This is because the Ward identity for the broken flavor symmetry implies a linear relation among the corresponding three operators in \Eq{trace}. 
Given any $S$, generator of $O(n)$, the Ward identity states
\beq
(S\cdot \lambda)_{ijk\ell}[{\cal O}_{ijkl}]+S_{ij}\left ([\phi_j E_i]+\partial_\mu [J_{ij}^\mu]\right )=0
\eeq
which added to \Eq{trace} implies  invariance under the  reparametrization
\beq
\be_{ijk\ell}\to \be_{ijk\ell}+(S\cdot \lambda)_{ijk\ell}\,,
\qquad
\Ga_{ij}\to \Ga_{ij}+S_{ij}\,,
\qquad 
N_{ij}\to N_{ij}+S_{ij}\,.
\eql{repar}
\eeq
Working in different renormalization schemes will in general lead to 
coefficients that differ by the above reparametrization.
This complete freedom was not discussed in \Ref{Osborn:1991gm}: 
in Eq.~(3.46) of that paper the ambiguity is limited to the 
case $S\propto \beta$.%
\footnote{In \Ref{BoazR} we will show how precisely the full ambiguity, 
with general $S$, arises when defining the RG flow in dimensional 
regularization. 
The basic point is that, given a theory renormalized with a well-defined pole subtraction procedure, there exists a family of possible choices of the RG flow coefficients, parametrized by \Eq{repar}, all describing the momentum evolution 
of the same correlators.}
The freedom in \Eq{repar} corresponds to the freedom in choosing 
the RG flow coefficients discussed for the case of limit cycles
in the Appendix.

A convenient way to ``fix the gauge" is to choose $S=-N$ so that the 
coefficient of the current in \Eq{trace} vanishes:
\beq
T = \frac{B_{ijk\ell}}{4!}[{\cal O}_{ijk\ell}]+  \De_{ij}[\phi_j E_i]+a_{ij}\Box[\phi_i\phi_j]
\eql{shorttrace}
\eeq
where $B_{ijk\ell}=\be_{ijk\ell}-(N\cdot \lambda)_{ijk\ell}$ and $\De_{ij}=\Ga_{ij}-N_{ij}$.
 According to the result proven in the previous section, 
 and further developed in the following sections, 
the only possible UV and IR asymptotics has $B=0$, corresponding to CFTs. 
In those asymptotic CFTs the eigenvalues of $\De_{ij}$ 
give  the scaling dimensions of the elementary fields. From  the  general theory of the unitary representations of $SO(4,2)$ \cite{Mack:1975je} these 
eigenvalues should all be real and $\geq 1$. 

This discussion also clarifies the significance of limit cycles with
$B = 0$, such as those discussed in \Ref{FGSta}.
In such theories there exists a ``gauge'' for $\be, N, \Gamma$ where 
$\be=0$ and $N=0$. 
This corresponds to a scheme where the couplings are constant on the RG flow. 
In a general gauge one would have $\be =(X\cdot \lambda)$ with $X$ a constant 
matrix: for these other choices the coupling would describe a cycle. 
It is however clear from the discussion that this cycling does not have
a reparameterization-invariant meaning.

In the appendix we consider theories with limit cycles that may or
may not have $B = 0$.
We show that these theories are also equivalent to theories
with fixed points due to the same ambiguity.

\subsection{Dilaton Couplings in Dimensional Regularization}
\label{sec:dimreg}

We will now make the above discussion more explicit by working in dimensional regularization
and defining composite operators by differentiation with respect to external sources, position dependent
couplings and flavor gauge fields. The discussion in this section is largely extracted from Ref.~\cite{Jack:1990eb}.
Moreover, since our final goal are the amplitudes between on-shell dilaton and on-shell fields, the equations of motion and the improvement terms will not play any role.

Consider the dimensionally regularized Lagrangian
\beq
{\cal L}_0 &= \sqrt{-g} \biggl[ 
\sfrac{1}{2} g^{\mu\nu} 
D_\mu \hat\Phi^i_0 D_\nu \hat\Phi^i_0 - \frac{\lambda_{0ijkl}}{4!} 
\hat\Phi^i_0\hat\Phi^j_0\hat\Phi^k_0\hat\Phi^l_0 
\nonumber\\
& \qquad\qquad\quad
{}+ N_{0mn}^{ijkl} g^{\mu\nu} D_\mu \la_{0ijkl} \Phi^m 
\!\! \stackrel{\leftrightarrow}{D}_\nu \!\! \Phi^n
+ \cdots \biggr] \,,
\eeq
where
\beq
D_\mu \hat{\Phi}^i = \d_\mu - (A_\mu)^i{}_j \hat{\Phi}^j
\eeq
is the flavor gauge covariant derivative.
Note that the kinetic term is chosen to be canonical and therefore
invariant under the flavor gauge group, which is $O(N)$ in this case.
Now $\la_0$ as well as $g_{\mu\nu}$ are functions of $x$, and the last 
term is required by renormalizability in the theory with position-dependent
couplings.
We are dropping terms that do not matter for getting the coupling of an on-shell dilaton to the QFT, 
including all curvature terms and higher derivatives of the coupling.
For a discussion of these terms, see \Ref{Jack:1990eb}.
Then for $g_{\mu\nu} = e^{-2 \tau} \eta_{\mu\nu}$ 
and defining $\Phi_0 = e^{(2-d) \tau/2} \hat\Phi_0$, this becomes
\beq
\label{eq:taucoup}
{\cal L}_0 =  \sfrac{1}{2}  D_\mu \Phi^i_0 D_\mu \Phi^i_0 
- e^{-\ep\tau} \frac{\lambda_{0ijkl}}{4!} \Phi^i_0\Phi^j_0\Phi^k_0\Phi^l_0 
+ N^{ijkl}_{0mn} D^\mu \lambda_{0ijkl} 
\Phi^m_0 \!\! \stackrel{\leftrightarrow}{D}_\mu \!\! \Phi^n_0   
\eeq
There is no $\partial\tau$ from the last term because of antisymmetry in $mn$.

Note that with the couplings defined as flavor spurions, the flavor
symmetry is exact, and so the flavor gauge fields do not need to be
renormalized.
$N_0$ can be chosen to have no finite part, so it is a sum of poles
in $1/\ep$.
We then write $\scr{L}_0 = \scr{L}_0^{(0)} + \scr{L}_0^{(2)} + \cdots$
where $\scr{L}_0^{(n)}$ is of order $\tau^n$.
The linear coupling of $\tau$ is then
\beq
\scr{L}_0^{(1)} &= 
\epsilon\tau \frac{\lambda_{0ijkl}}{4!} \Phi^i_0\Phi^j_0\Phi^k_0\Phi^l_0 
\nonumber\\
&= -\epsilon\tau \lambda_{0ijkl} \frac{\de{S}_0}{\de \lambda_{0ijkl} }+ \epsilon\tau \lambda_{0ijkl} N^{ijkl}_{0mn}  \partial_\mu\Phi^m_0 \partial_\mu\Phi^n_0 \nonumber\\
&=  -\epsilon\tau \lambda_{0ijkl} \frac{\partial \lambda_{i'j'k'l'}}{\partial \lambda_{0ijkl} }\frac{\de{S}_0}{\de \lambda_{i'j'k'l'} } 
+ \epsilon\tau \lambda_{0ijkl} N^{ijkl}_{0mn} 
 \partial_\mu\frac{\de{S}_0}{\de \bar A_{mn}^\mu }  \,,
\eql{linear}
\eeq
where $\bar{A}_{\mu mn}$ is defined by
\beq
\bar{A}_{\mu mn} = A_{\mu mn} + N_{0ijkl}^{mn} (A_\mu \la_0)^{ijkl}.
\label{Ashift}
\eeq
The second term comes from the covariant derivative acting on $\la_0$
in the Lagrangian.
We can simplify the first term in \Eq{linear}
using the dimensional renormalization relation
\beq
\epsilon \lambda_{0ijkl} + \frac{\partial \lambda_{0ijkl} }{\partial \lambda_{i'j'k'l'}}\be_{i'j'k'l'} = 0\,,
\eeq
and it is then explicitly finite.  
Since the total $\tau$ coupling is renormalized, the second term 
must also be finite. 
By expressing it in terms of  $\de/\de A_{mn}^\mu$, which is also 
finite by construction, the only surviving term in the coefficient must be 
the one proportional to $\epsilon^0$ term. 
Because $N_0$ is a series of pure poles, the $\epsilon^0$ term is just given 
by the residue of the single-pole term in $N_0$.%
\footnote{Note that, via eq.~(\ref{Ashift}), the $1/\ep$ pole in $N_0$ also gives a contribution
to the anomalous dimension of the flavor current.}
Therefore, the linear coupling of $\tau$ is 
\beq
\scr{L}_0^{(1)} =  \tau \be_{i'j'k'l'} \frac{\de{S}_0}{\de \lambda_{i'j'k'l'} } 
+ \tau \lambda_{ijkl} N^{ijkl}_{1mn} 
 \partial_\mu\frac{\de{S}_0}{\de A_{mn}^\mu }  
\eeq
where $N_1$ is the residue of the $1/\ep$ pole in $N_0$, which is a finite
quantity.
The second term can be simplified via the equation of 
motion~\cite{Jack:1990eb}
(in the notation of that paper $S = \lambda N_1$).
We then obtain
\beq
\scr{L}_0^{(1)} = 
 \tau B_{i'j'k'l'} \frac{\de{\cal L}_0}{\de \lambda_{i'j'k'l'} } \,.
\eeq

Now consider the couplings of order $\tau^2$.  
From \Eq{taucoup}, 
\beq
\scr{L}_0^{(2)} = \sfrac 12 \ep \scr{L}_0^{(1)}.
\eeq
However, since $\scr{L}_0^{(1)}$ is finite, the quadratic coupling must 
vanish at $d=4$.  
A local $\tau^2$ coupling does arise from a contact term of 
order~$B \partial_\lambda B$ appearing  from double insertion of the 
linear coupling to $\tau$~\cite{BoazR}. 
However, this term is parametrically suppressed compared to the
$O(B)$ term arising from the nonlinear 
relation between $\tau$ and $\varphi$, which was used in the
argument of \S\ref{sec:j=0}.

\subsection{Constraints from Wess-Zumino Consistency}
\label{sec:wzcon}
The discussion in this section is largely extracted from \Ref{Osborn:1991gm}.
A flow equation of the same form as \Eq{alrun} can be obtained 
by application of the WZ consistency relations.  
Consider a renormalized generating functional 
$W[g_{\mu\nu}, \la_A, A^{\mu a}]$.  
Here $\la_A$ are a complete set of couplings for dimension-4 scalars and 
$A^{\mu a}$ are a complete set of flavor gauge fields,
that is, couplings for dimension-3 vector operators.
The couplings $\la_A$ as well as $g_{\mu\nu}$ and $A_{\mu a}$ are
allowed to depend on $x$.  
Various dimension-2 scalars are also needed, but do not enter into the 
following discussion.
The couplings of $A_{\mu a}$ are restricted by background gauge invariance 
and so it  appears in covariant derivatives and field strengths, 
\eg~$D_\mu \la_A = \partial_\mu \la_A + A_{\mu a} (T_{a})_A{}^B \la_B$,
where $T_a$ are the generators of the flavor group.

The Weyl variation of $W$ can be written in terms of a sum of all local terms,
$\de_x W = D(x)$, where $D(x)$ includes all possible dimension-4
functions of the sources, including gradients of couplings and curvatures.
The Weyl variation operator is
\beq
\de_x = \frac{\de}{\de \tau(x)} 
- \be_A \frac{\de}{\de \la_A(x)}  
- \rho^A_a D_\mu \la_A \frac{\de}{\de A_{\mu a}(x)} 
- D_\mu \left( S_a \frac{\de}{\de A_{\mu a}(x)} \right)
\eeq
Using background gauge invariance, we can collect the last term into the 
others:
\beq
\de_x = \frac{\de}{\de \tau(x)} 
- B_A \frac{\de}{\de \la_A(x)}  
- P^A_a D_\mu \la_A
\frac{\de}{\de A_{\mu a}(x)} ,
\eeq
with $B_A = \be_A - (S \la)_A$ and 
$P^A_a (D_\mu \la)_A = \rho^A_a (D_\mu \la)_A + (D_\mu S)_a$.  
That is, $B$ is the total Weyl variation of the dimension-4 couplings, 
and $P$ is the total Weyl derivative of the vector couplings.  
Thus the condition for conformal invariance is $B=0$.

The Wess-Zumino consistency condition is
\beq{}
[\de_x ,\de_y]W = \de_x D(y) - \de_y D(x).
\eeq
To derive the flow equation, we only need terms proportional to the 
Einstein tensor, which arise from
\beq
D(x) = \be_{(b)} E_4 
+ \sfrac{1}{2} G^{\mu\nu} \chi_{(\la)}^{AB} D_\mu \la_A D_\nu \la_B 
+ G^{\mu\nu} D_\mu(w^A D_\nu \la_A) + \ldots\,.
\eeq
Picking out terms proportional to 
$G^{\mu\nu} D_\mu \la_A \de'(x-y)$, the consistency relation becomes 
\beq
8 \frac{\d \be_{(b)}}{\d\la_A} = \chi^{AB}_{(\la)} B_B
- \frac{\d w^A}{\d\la_B} B_B 
- \frac{\d B_B}{\d\la_A} w^B + P^A_a (T_a \la)_B w^B \,.
\eeq

Now, contracting with $B_A$, defining 
$\tilde B_{(b)} = \be_{(b)} + \frac 18 w^A B_A$, 
and using $B_A P^A = 0$ (from the $\de/\de A$ term in the 
Wess-Zumino condition) gives~\cite{Osborn:1991gm,FGSJO}
\beq
8B^A \partial_A \tilde B_{(b)} = \chi^{AB}_{(\la)} B_A  B_B\,.
\eeq
This gives the flow equation
\beq[bbtflow]
\mu\frac{d \tilde B_{(b)}}{d\mu}  
= B_A \frac{\d \tilde{B}_{(b)}}{\d\la_A}
= \be_A \frac{\d \tilde{B}_{(b)}}{\d\la_A}
= \sfrac 18 \chi^{AB}_{(\la)} B_A B_B.
\eeq
(Note that $B$ and $\be$ give the same flow here because $\tilde B_{(b)}$ 
is invariant under field rotations).

In perturbation theory, $\chi^{AB}_{(\la)}$ is positive, and the parallel between 
the flow~\Eq{bbtflow} of $\tilde B_{(b)}$ and the flow~\Eq{alrun} of 
$\bar\alpha$ is evident.  
In particular, each is stationary only at conformal points, and so any 
perturbative flow must approach a conformal theory in the UV and the IR.%
\footnote{As will be shown in Ref.~\cite{BoazR}, 
$\chi^{AB}_{(\la)}$ is indeed positive definite for perturbations
around any conformal fixed point.}

\section{Scale Versus Conformal Invariance}
\label{sec:SwoC}
In this section, we consider in more generality
the question of whether scale invariance
implies conformal invariance, without assuming the validity of perturbation 
theory.
In 2D there is a rigorous argument
that scale invariance implies conformal invariance \cite{Polch}
based on the Zamolodchikov $c$-theorem \cite{Zamol}.
Given that we now have a (non-perturbative) proof of the $a$-theorem
in 4D, it is natural to ask whether we can give a similar proof
for 4D theories.  
This was shown at the classical level in Ref.~\cite{CCJ}, 
but of course quantum effects play an essential role. 

The results of \S\ref{sec:pertRG}
already show that scale invariance implies conformal invariance for
weakly-coupled flows.  
In particular, for nonconformal scale invariant theories (SFTs) the dilaton 
does not decouple; see also the recent discussion~\cite{Nakayama:2011wq}.
For non-perturbative theories, we will show that scale invariance
implies conformal invariance subject to a plausible technical assumption.
Various aspects of our argument will also be checked in perturbative
examples that have some but not all of the features of unitary SFTs.

\subsection{Generalities}
Consider a possible theory that is scale invariant but not
conformally invariant, a SFT.
It was shown by Wess \cite{Wess}
that the most general conserved scale current has the form
\beq[scalecurrent]
S^\mu = T^\mu\!_\nu x^\nu + V^\mu,
\eeq
where $T^\mu\!_\mu$ is the stress-energy tensor
and $V_\mu$ is called the virial current.
(We will not consider scale invariant theories that do not have a local
scale current.)
Conservation of the scale current then implies
\beq
0 = \d_\mu S^\mu = T + \d_\mu V^\mu.
\eeq
The theory is conformal if $T \equiv 0$,
so we see that we can get a nontrivial SFT only if
$\d_\mu V^\mu \ne 0$.

We can write the Ward identities for scale invariance in a 
convenient form by introducing a source $C_\mu$ for the virial
current as well as using $g_{\mu\nu}$ as a source for $T^{\mu\nu}$.
For local scale transformations generated by $\si(x)$ we then have
\beq[SFTWard]
W_{\rm SFT}[e^{2\si} g_{\mu\nu}, C_\mu + \d_\mu\si]
= W_{\rm SFT}[g_{\mu\nu}, C_\mu]
+ S_{\rm WZ}[\si; g_{\mu\nu}, C_\mu],
\eeq
where $S_{\rm WZ}$ is an anomaly term.
The anomaly term is local and must satisfy the Wess-Zumino
consistency conditions, which enforce that local scale transformations
are Abelian.
In addition, the WZ term cannot contain any dimensionful parameters.

The most general WZ term satisfying these constraints can be
readily found.
It includes the $a$ and $c$ terms from conformal field theories,
which do not depend on $C_\mu$.
There are only two additional allowed terms
\beq[newanomaly]
\De S_{\rm WZ} = 
\myint d^4 x\, \sqrt{-g} \, \si \left[ 
e\, \scr{O}_e + f\, \scr{O}_f \right]
\eeq
where
\beq
\!\!\!\!\!\!\!
\eql{eanomaly}
\scr{O}_e &= 
\sfrac 1{12} R^2(g)
+ R(g) \nabla \cdot C
- R(g) C^2 
+ 3 (\nabla \cdot C)^2
-6  C^2 \nabla \cdot C
+ 3 C^4,
\\
\scr{O}_f &= C^{\mu\nu} C_{\mu\nu},
\eeq
where $C_{\mu\nu} = \d_\mu C_\nu - \d_\nu C_\mu$.
The $e$ term will play an important role in our arguments below.

\subsection{Non-perturbative Argument
\label{sec:NPSFT}}
We again consider the amplitude $A(s)$ defined above.
That is, we consider background fields
\beq
\hat{g}_{\mu\nu} = \left( 1 + \frac{\vph}{f} \right)^2 \eta_{\mu\nu},
\qquad
C_\mu \equiv 0,
\eeq
and define ``dilaton amplitudes'' by differentiating
$W[\hat{g}_{\mu\nu}, 0]$ with respect to $\vph$,
taking the forward limit,
and imposing the ``on-shell'' condition $\Box\vph = 0$.

We can find the exact form of this amplitude in an SFT using the
Ward identity \Eq{SFTWard} with $g_{\mu\nu} = \hat{g}_{\mu\nu}$,
$C_\mu = 0$, and $\si = \mbox{constant}$.
The anomaly term in \Eq{SFTWard}
does not contribute to the transformation
of $A(s)$:
the $a$ term vanishes (for $\si = \mbox{constant}$)
because $E_4(g)$ is a total derivative,
the $c$ term vanishes for conformally flat backgrounds,
the $e$ and $f$ terms vanishes because of the on-shell condition $R(g)=0$ and because we are considering backgrounds
with $C_\mu = 0$.
The fact that there is no anomalous scaling of the amplitude
is equivalent to the statement that there is no counterterm
for this amplitude that can have a logarithmic dependence
on the cutoff.

Because there is no anomalous contribution to the amplitude
$A(s)$, it has the exact form dictated by \naive\ scale
invariance, namely
\beq[scaleinvsigma]
A(s) = \frac{\al s^2}{f^4},
\eeq
with $\al = \mbox{constant}$.
(Here we are assuming that the vacuum is scale invariant,
so this does not hold for nonlinearly realized scale invariance.
An example of this kind will be discussed in \S\ref{sec:examples} below.)
This has no imaginary part, which has strong implications
in a unitary theory.
Unitarity implies that
the imaginary part of $A(s)$ is positive, 
and in fact can be written as a sum over a complete set of state
states, with each state giving a positive contribution.
Therefore, the only way we can have a vanishing imaginary 
part is for each term
to vanish individually.
We must therefore have
\beq[statezerocond]
\scr{O}(p_1, p_2) \ket{0} = 0,
\eeq
where 
\beq
\scr{O}(p_1, p_2) =  T(p_1) T(p_2) 
+ T(p_1 + p_2).
\eeq
(Here the trace of the stress tensor is defined as in \Eq{goodT}
above.)
If \Eq{statezerocond} were to hold for arbitrary momenta $p_{1,2}$
then we could immediately conclude that $\scr{O}(p_1, p_2) \equiv 0$
as an operator statement.
In position space, this is the statement that
\beq
\mbox{T} \bigl\{ T(x_1) T(x_2) \bigr\}
= - \de^4(x_1 - x_2) T(x_1),
\eeq
which can only hold in a theory where $T$ is a trivial
operator.
However, our arguments hold only for on-shell dilaton amplitudes
with $p_{1,2}^2 = 0$.
We cannot extend our arguments to off-shell amplitudes
because then the $e$ anomaly term \Eq{eanomaly} contributes
to the scale transformation of $A(s)$, allowing a log
term in $A(s)$.
This corresponds to a logarithmic renormalization of an
$R^2$ counterterm.%
\footnote{As discussed in \S\ref{sec:Dele2}, a logarithmic renormalization
of $R^2$ arises in CFTs deformed by a term $\scr{R} \scr{O}$ where 
$\scr{O}$ has dimension 2 (\eg\ $\scr{O} = \Phi^2$ in a theory of a
free scalar).
This is not surprising since this is an explicit marginal breaking of
conformal invariance.}

It is interesting to contrast this situation with the
$d=2$ argument~\cite{Polch, Zamol}.  
In that case, the vanishing of the $T$ 2-point function
immediately implies that $T$ must vanish (with no on-shell
condition).

One scenario where the amplitude $A(s)$ has no 
imaginary part in a SFT with a nontrivial operator $T$
is to have a factorized $T$ amplitude:
\beq[Tdisconnected]
\!\!\!\!\!\!
\bra{0} & {\rm T} \bigl\{ T(x_1) T(x_2) T(x_3) T(x_4) \bigr\} \ket{0}
\nonumber\\
&\qquad\quad
 = \frac{\mbox{constant}}{(x_1 - x_2)^8 (x_3 - x_4)^8} 
+ \mbox{\ permutations} + \mbox{contact terms} \,.
\eeq 
The nonlocal part of this amplitude contributes only to
forward scattering, and therefore does not contribute to $\Im A$.
Then $\Im A$ is completely local, and this local term
can vanish.
(Note that large-$N$ theories do not provide the necessary factorized form.
The leading contribution in the large-$N$ limit is disconnected,
connected amplitudes remain at subleading order in the $1/N$ expansion.)
The disconnected amplitude \Eq{Tdisconnected} implies a $T T$ 
operator product expansion (OPE)
that contains only the unit operator and nontrivial operators starting
at dimension 8.
Even if the theory satisfies all this, $C = 0$ requires a cancellation
between the (presumably infinitely many) operators appearing in the $TT$
OPE and the $TT$ contribution in \Eq{AintermsofT}.
In a CFT we could conclude that $T$ appears in the $TT$ OPE
from the symmetry of the OPE coefficients, 
but it is not clear that this is so in an SFT, 
as the position dependence of the three-point function is undetermined.
We regard this exception as implausible, 
and likely to be ruled out in the future.

\subsection{Examples
\label{sec:examples}}
For a nonunitary theory, a nonzero $T$ can still give rise to a vanishing
$C$ by cancellation between positive and negative contributions.  
Our argument, that divergent renormalization of $a$ is inconsistent, 
requires that such cancellation occur in any nonunitary SFT.  
A massless vector field without gauge invariance provides a such a theory, 
in any dimension.  
This was first noted by Coleman and Jackiw~\cite{ColeJack} in $d=4$.  
It was studied in detail in $d=2$ by Riva and Cardy (RC) \cite{RC},
who also observed that it has a Euclidean interpretation as the 
theory of elasticity.   

The action is written in terms of a displacement field~$u_\mu(x)$.
Following RC we start with a Euclidean action, 
which is the physically relevant signature for elasticity,
\beq[elastic]
S = \int d^4x \sqrt{-g}\left( \frac{1}{4} F_{\mu\nu} F^{\mu\nu} + \frac{h}{2} (\nabla_\mu u^\mu)^2 \right) \,.
\eeq
Here $F_{\mu\nu} = \partial_\mu u_\nu -  \partial_\nu u_\mu$.  
(In the notation of RC, $g=1$ and $k = h-2$.)

Scale invariance is manifest, with $u_\mu$ having dimension~1.  
Consider now the metric $g_{\mu\nu} = e^{-2\tau} \de_{\mu\nu}$.  The Maxwell-like term is conformally invariant in $d=4$.  Using $\nabla_\mu u^\mu = \sqrt g^{-1} \partial_\mu (\sqrt g g^{\mu\nu} u_\nu) = e^{4\tau} \partial_\mu (-e^{2\tau} u_\mu)$,  the coupling the dilaton is
\beq
2 h \,\partial_\mu u^\mu\, u^\nu {} \partial_\nu\varphi -  2 h \,\partial_\mu u^\mu\, u^\nu \varphi \partial_\nu\varphi  + 2h (u^\mu {} \partial_\mu\varphi)^2 + O(\varphi^3)
\,, \label{vertex}
\eeq
and so the theory is not conformally invariant.  We can regulate this theory using Pauli-Villars in the background $\hat g_{\mu\nu}$.  Our arguments then imply that there should be no running of $a$.  This is true, but it is nontrivial, and provides a test of our reasoning.

If we rotate to Lorentzian signature, these couplings give a  $\varphi \varphi \to uu$ scattering amplitude, which has a contact piece from the $\varphi^2$ terms in the interaction (\ref{vertex}) and pole and contact pieces from second order in the $\varphi$ term.   If the total cross section obtained by `squaring' and integrating over phase space were nonzero, it would be proportional to $s^2$ from the scaling of the vertices, and the dispersive argument in Fig.~1 would imply a logarithmic divergence of $a$; we have argued this to be inconsistent.  In fact, direct calculation shows that there is no logarithmic divergence, and so no imaginary part in the forward amplitude.  Individual graphs are nontrivial functions of $h$, but the sum vanishes identically.  This cancellation provides a satisfying check of the general argument.  
Of course, the vanishing of the total cross section is possible only 
because of the nonunitarity of the theory. 
In a unitary theory the {\it differential} cross section would have to vanish identically.

The calculation of the logarithmic divergence is uninstructive, but we can give an independent indirect argument for it, exploiting the resemblance between $u_\mu$ and a gauge field.  First, we can write the theory in a gauge invariant way by introducing a Goldstone field $\pi$,
\beq
S = \int d^4x \sqrt{-g}\left( \frac{1}{4} F_{\mu\nu} F^{\mu\nu} +  \left[ \nabla^\mu ( \partial_\mu \pi +  h^{1/2} u^\mu) \right]^2 \right) \,.
\eeq
In the gauge $\pi = 0$ this reduces to the earlier action.  Now, at $h = 0$ this becomes
\beq
S = \int d^4x \sqrt{-g}\left( \frac{1}{4} F_{\mu\nu} F^{\mu\nu} +  (\Box \pi)^2 \right) \,.
\eeq
Both terms are now conformally invariant and so the on-shell dilaton coupling vanishes.  In particular, by adding improvement terms $\Box \pi \to (\Box + R/6) \pi$ we can make the action fully Weyl-invariant.

This implies the vanishing of the cross section at $h = 0$.  To extend this, we look at the action in a different way.  The $h(\nabla_\mu u^\mu)^2 /2$ term can be interpreted as a covariant gauge-fixing term.  We might then conclude that the dilaton decouples due to the conformal invariance of the pure Maxwell theory.  However, we did not include the associated ghost determinant, which will have a nontrivial dependence on the metric.  The ghost action is independent of $h$, and so we can argue from this that the renormalization of $a$ is independent of $h$, and so vanishes for all $h$.

Another potential counterexample noted in the classic 
literature~\cite{CCJ} is the scalar Lagrangian
\beq[CCJex]
{\cal L} = \sfrac 12 \d^\mu \hat\Phi \d_\mu \hat\Phi 
+ \frac{(\d^\mu \hat\Phi \d_\mu \hat\Phi)^2}{\hat\Phi^4} \,.
\eeq
This Lagrangian is manifestly scale invariant but not conformally invariant.  
It is nonrenormalizable, but in a state with $\avg{\hat\Phi} \ne 0$ we can still consider it as an effective field theory below the scale $\avg{\hat\Phi}$,
with scale invariance nonlinearly realized in 
an expansion in inverse powers of $\avg{\hat\Phi}$.
There is no symmetry forbidding a $\hat{\Phi}^4$ term, which would
forces the ground state to be at $\avg{\hat\Phi} = 0$, 
but this term can be tuned away.
There are also an infinite number of additional terms with higher
powers of derivatives that give higher order corrections
in the low-energy expansion.
These do not affect the arguments below.

As noted below \Eq{scaleinvsigma}, our general argument ruling out
SFTs assumes that scale invariance is linearly realized, and so
this theory does not conflict with those arguments.
However, the argument that $a$ is not renormalized in a SFT
applies to this theory, so we can check it.
In flat spacetime, it is natural to expand 
\beq
\hat\Phi = v + \hat{\pi}.
\eeq
Setting $\hat{g}_{\mu\nu} = \Om^2\eta_{\mu\nu}$
and
\beq
\hat\pi = \Om^{-1} \pi,
\eeq
the Lagrangian becomes
\beq
\scr{L} = \sfrac 12 (\d \pi)^2 +
\frac{[\d (\Om^{-1}\pi)]^4}{(v + \Om^{-1} \pi)^4}.
\eeq
where we set $\Box\Om = 0$.
Note that there is no mixing between the background dilaton
and the dynamical dilaton field $\pi$.
All interactions of the dilaton involve inverse powers of the symmetry 
breaking scale $v$, giving a cross section that scales as
\beq
\si(\vph\vph \to \pi\pi) \sim \frac{s^5}{v^8 f^4}.
\eeq
Higher-order interactions will give additional terms with
additional positive powers of $s/v^2$.
These give power law divergences in the dispersion integral, 
which just cancel the negative powers of $v$ from the cross section:
there is no logarithmic dependence on the UV cutoff,
consistent with our general arguments.

\section{Conclusions}
We have shown that the method introduced in \Refs{KS,Ka}, 
using properties of the $a$-anomaly 
to show irreversibility in the flow between 4D CFT's, 
can be extended to exclude a wide variety of other 4D flows.  
The key observation is that the total flow of the scattering amplitude 
$\alpha(s)$ defined in \Eq{alphadefn} must be finite.  
We can then show rigorously that the only IR or UV asymptotics that can 
be described in
perturbation theory is that of a conformal field theory.
This excludes theories with scale without conformal invariance,
such as those proposed in \Ref{FGSexample}.
We can extend this argument beyond perturbation theory to rule out
scale without conformal invariance in theories that can be deformed
to a CFT in the UV or IR at an adjustable scale.
We can rule out general non-perturbative theories with scale but
not conformal invariance subject to a technical assumption:
that vanishing of the imaginary part of the
amplitude \Eq{AintermsofT} implies vanishing of $T^\mu\!_\mu$.  
This is quite plausible, as the imaginary part of the amplitude 
receives contributions of the same sign from all intermediate states, 
so all of these must vanish.

We have tested our argument against various possible counterexamples.  
The non-unitary theory of Riva and Cardy
evades the argument by cancellation of positive and negative contributions, 
but confirms our key observation that the renormalization of $\alpha$ must 
be finite.  
Theories with nonlinearly realized scale invariance are not ruled
out by our arguments, but the argument that $\alpha$ has no logarithmic
dependence on the cutoff can be checked in these theories.
Our results predict that the limit cycles discussed in 
\Ref{FGSexample} are conformal, as has been recognized by
the authors \cite{FGSta}.

Monotonicity of the RG flow has not been established aside from 
2 and 4 dimensions, 
although various ideas are being explored.
The apparent lesson from the known cases is that monotonicity arguments
can be extended to exclude SFT's and other exotic RG flows, 
but that these are not immediate corollaries; rather, they use the same 
machinery in new ways.  

Finally, any discussion of RG flows will have a geometric analog via holography. 
\Refs{Nakayama} investigate possible holographic realizations of theories
with scale but not conformal invariance. 
These theories violate the
null energy condition, strongly suggesting that they are unphysical. 
On the other hand, our arguments rely on the
renormalization properties of the theory in some background state, and
do not appear to require the stability of this state. 
The consistency of the holographic theories with scale and not 
conformal invariance and their relation to our work is an interesting 
subject for future investigation.

\section*{Acknowledgements}
We thank D. Shih for collaboration in the early stages of this work,
and S. Dubovsky, J.-F. Fortin,
B. Grinstein, B. Keren-Zur, J. Maldacena, H. Osborn, D. Pappadopulo, 
I. Rothstein, S. Rychkov, A. Stergiou, R. Sundrum, and A. Zaffaroni for discussions.
MAL and RR thank the KITP Santa Barbara for hospitality
during the initial stages of this work.  
This research was supported in part by the NSF under grants PHY07-57035 and PHY11-25915, by the DOE under grant
DE-FG02-91-ER40674 and by the Swiss National Science Foundation under grant 200021-125237.

\appendix{Appendix: Callan-Symanzik Equations, Flavor, and Limit Cycles}
In this appendix we show that limit cycles are completely equivalent to
fixed points due to the inherent ambiguity in renormalizing theories
with flavor.
We will actually prove a slightly stronger result, namely that any
RG flow that is equivalent to a 
scale-dependent flavor rotation is equivalent to a fixed point.
The renormalization ambiguity described here for limit cycles is
a special case of the one discussed in \S\ref{sec:Toperator},
and the precise connection will be made explicit in \Ref{BoazR}.

To illustrate our ideas, we consider $\Phi^4$ theory with $N$ scalar
fields, with renormalized Lagrangian
\beq[LRapp]
\scr{L}_R = \sfrac 12 \d\Phi^i \d\Phi^i
- \frac{\la_{ijkl}}{4!} 
\Phi^i \Phi^j \Phi^k \Phi^l,
\eeq
The renormalized Lagrangian \Eq{LRapp} contains all the information that
is needed to compute the amplitudes of the theory.
The counterterms, and hence the bare couplings, are determined order-by-order
in perturbation theory by requiring the cancelation of the $1/\ep$ divergences 
that appear.

We now turn to the Callan-Symanzik equations of the theory, which
we will see are ambiguous in theories with flavor.
The Callan-Symanzik equations state that correlation
functions of the fields are independent of $\mu$ up to a 
$\mu$-dependent rescaling of the fields.
We can express this as the requirement that the correlation
functions of rescaled fields $\hat\Phi$ are independent of $\mu$.
We write
\beq
\Phi^i = \xi^i{}_a \hat{\Phi}^a.
\eeq
Note that we are free to redefine the fields $\hat{\Phi}^a$
by a $GL(N)$ transformation acting on the $a$ index,
corresponding to an arbitrary field redefinition that is linear
in the fields.
It is conventional to choose $\xi^a{}_i$ to be symmetric, but
there is nothing that forbids a choice where $\xi^i{}_a$ 
has an antisymmetric part.
We will see below that this freedom allows us to rewrite limit
cycles as fixed point.
The Callan-Symanzik equation is then
\beq\!\!\!\!\!\!\!
&\left( \frac{\d}{\d t}
+ \be_{ijkl} \frac{\d}{\d\la_{ijkl}} \right) \avg{\Phi^{i_1} \cdots \Phi^{i_n}}
\nonumber\\
&\qquad\qquad{}
= \ga^{i_1}{}_{k} \avg{\Phi^{k} \Phi^{i_2} \cdots \Phi^{i_n}}
+ \cdots + \ga^{i_n}{}_k \avg{\Phi^{i_1} \cdots \Phi^{i_{n-1}} \Phi^k}\,.
\eeq
where
\beq
\frac{d}{dt} \la_{ijkl} &= \be_{ijkl}(\la),
\\
\eql{wavfuncngamma}
\frac{d}{dt} \xi^i{}_a &= \ga^i{}_j(\la) \xi^j{}_a.
\eeq
The form of \Eq{wavfuncngamma} and the fact that $\be$ and $\ga$ do not depend
on $\xi$ can be understood from covariance under $GL(N)$.

We now consider RG flows in which RG flow
is equivalent to a flavor rotation, \ie
\beq[flavorflow]
\la(t) = R(t) * \bar\la\,,
\eeq
where $R(t)$ is a scale-dependent $O(N)$ transformation
and $\bar\la$ is a scale-independent renormalized coupling.
We are using an abstract notation where $*$ denotes the $O(N)$
action in the appropriate representation.
For example, \Eq{flavorflow} is short for
\beq
\la_{ijkl}(t) = \bar\la_{i'j'k'l'} (R^{-1}(t))^{i'}{}_i (R^{-1}(t))^{j'}{}_j
(R^{-1}(t))^{k'}{}_k (R^{-1}(t))^{l'}{}_l.
\eeq

If we make the redefinition of renormalized fields
\beq[Rfieldredefn]
\Phi^i = R^i{}_j(t) \Phi'^j,
\eeq
the renormalized Lagrangian becomes
\beq
\scr{L}_R = \sfrac 12 \d \Phi'^i \d \Phi'^i - \frac{\bar{\la}_{ijkl}}{4!}
\Phi'^i \Phi'^j \Phi'^k \Phi'^l.
\eeq
That is, the theory is equivalent to a fixed point theory.%
\footnote{In more general flows it is also 
natural to make the field redefinition
\Eq{Rfieldredefn} where $R$ depends on $t$ via the renormalized couplings
$\la(t)$.
This changes the beta function, but it cannot relate a fixed point
theory to one with nonzero beta functions.}
We emphasize that \Eq{Rfieldredefn} is a completely finite
field redefinition, and therefore there is no need to
reconsider the renormalization of the theory in terms of 
the new fields.

In fact, we can determine the explicit form of $R(t)$ as follows.
Requiring that \Eq{flavorflow} solves the RG
equations gives
\beq
\frac{d}{dt} \left( R(t) * \bar\la \right)
= \be(R(t) * \bar\la) = R(t) * \be(\bar\la),
\eeq
where we have used the flavor covariance of the $\be$ function
in the last step.
This just means that the only violation of flavor symmetry
comes from the couplings, which can be taken to be spurions
for the flavor symmetry.
We therefore have
\beq
\left[ R^{-1}(t) \frac{d}{dt} R(t) \right] * \bar\la = \be(\bar\la).
\eeq
Since the \rhs\ is independent of $t$, the left-hand-side must be
as well, which means that the scale-dependent $O(N)$ transformation
must have the form
\beq
R(t) = e^{-t X}
\eeq
for some fixed flavor generator $X$
constructed from the couplings $\bar\la$.
This is the form of the RG cycles considered by \Refs{FGSexample,FGSta}.

We can similarly analyze the RG equation for the
wavefunction factors, \Eq{wavfuncngamma}.
Covariance of the anomalous dimension function $\ga$ implies that
\beq
\frac{d}{dt} \xi^i{}_a
= (e^{-t X})^i{}_{i'} \ga^{i'}{}_{j'}(\bar{\la}) 
(e^{t X})^{j'}{}_j \xi^j{}_a.
\eeq
This is solved by 
\beq
\xi(t) = e^{-t X} e^{t (X + \bar{\ga})},
\eeq
where $\bar{\ga} = \ga(\bar{\la})$.
This solution is not unique because we can multiply on the
right by an arbitrary $t$-independent linear transformation.
This solution therefore makes a
canonical identification between $a$ and $i$ indices.

In terms of the new fields $\Phi'$ given by \Eq{Rfieldredefn}, we have
\beq
\xi'\ = R^{-1} \xi  = 
e^{t (X + \bar{\ga})}\,.
\eeq
The renormalized Lagrangian is therefore invariant under
scale transformations  of the form
\beq
\Phi'(x) \mapsto e^{\De  t} \Phi'(e^t x).
\eeq
Here
\beq
\De = \De_0 + \bar{\ga}+X
\eeq
where $\De_0$ is the canonical dimension
of the fields ($1$ for scalars $\frac 32$ for fermions).
Note that gauge fields do not have independent wavefunction
renormalization factors.
For example, the 2-point function of these fields is given by
\beq
\avg{\Phi'^i(x) \Phi'^j(0)}
= \frac{1}{x^\De} C \left( \frac{1}{x^\De} \right)^T.
\eeq
where $C^{ij} = C^{ji}$ are constants.
We see that in terms of the fields $\Phi'$ the scale
invariance is manifest. 
According to the main result of this paper, scale invariance should  extend to the full conformal group. The elementary fields $\Phi_i$ must then correspond to a set of primary scalars whose scaling dimensions are determined by the eigenvalues of $\Delta$. The general result \cite{Mack:1975je} for the spectrum of the dilation operator in CFTs 
imposes then a further constrain on $\Delta$: even though it is in general not symmetric it should be diagonalizable with  real eigenvalues $\geq 1$.

%

%

\end{document}